\documentclass[12pt]{article}

\usepackage{amsmath}
\usepackage{amssymb}
\usepackage{amsthm}
\usepackage{mathrsfs}
\usepackage[T1]{fontenc}
\usepackage{enumerate}
\usepackage{color}

\def\eqn{equation}

\def\comrel{comutation relation}
\def\tfn{transformation}

\def\sm{sigma model}

\def\dd{Drinfel'd double}
\def\vbe{beta function equations}
\def\4diml{four-dimensional}
\def\bkg{background}
\def\wrt{with respect to}
\def\-1{^{-1}}
\def\half{\frac{1}{2}}
\def\coor{coordinate}

\def\e{{e}}

\def\cd{{\mathfrak d}}
\def\cg{{\mathfrak g}}
\def\tcg{\tilde{\mathfrak g}}
\def\hcg{\hat{\mathfrak g}}
\def\bcg{\bar{\mathfrak g}}

\def\wt{\tilde}
\def\wh{\widehat}
\def\wwt{\widetilde}
\def\sm{sigma model}
\def\pltp{Poisson--Lie T-pluralit}
\def\pltd{Poisson--Lie dualit}

\def\natd{(non-)Abelian T-dualit}
\def\sugra{Generalized Supergravity Equation}

\def\cf{{\mathcal {F}}}

\newcommand{\unit}{\mathbf{1}}
\newcommand{\nul}{\mathbf{0}}
\newcommand{\A}{\mathscr{A}}
\newcommand{\B}{\mathscr{B}}
\newcommand{\M}{\mathscr{M}}
\newcommand{\D}{\mathscr{D}}
\newcommand{\G}{\mathscr{G}}
\newcommand{\tG}{\widetilde{\mathscr{G}}}
\newcommand{\hG}{\widehat{\mathscr{G}}}
\newcommand{\bG}{\bar{\mathscr{G}}}

\newcommand{\hJ}{\widehat{\mathcal{J}}}
\newcommand{\bJ}{\bar{\mathcal{J}}}

\newcommand{\N}{\mathscr{N}}

\newcommand{\PL}{Poisson--Lie}

\newcommand{\wb}{\bar}

\title{Poisson--Lie plurals of Bianchi cosmologies and Generalized Supergravity Equations}
\author{Ladislav Hlavat\'y\footnote{hlavaty@fjfi.cvut.cz}
\\ {\em Faculty of Nuclear Sciences and Physical Engineering,}
\\ {\em Czech Technical University in Prague,}
\\ {\em Czech Republic}
\and
Ivo Petr\footnote{ivo.petr@fit.cvut.cz}
\\ {\em Faculty of Information Technology,}
\\ {\em Czech Technical University in Prague,}
\\ {\em Czech Republic}}

\begin{document}
\maketitle

\begin{abstract}
Poisson--Lie T-duality and plurality are important solution generating techniques in string theory and (generalized) supergravity. Since duality/plurality does not preserve conformal invariance, the usual beta function equations are replaced by Generalized Supergravity Equations containing vector $\mathcal{J}$. In this paper we apply Poisson--Lie T-plurality on Bianchi cosmologies. We present a formula for the vector $\mathcal{J}$ as well as transformation rule for dilaton, and show that plural backgrounds together with this dilaton and $\mathcal{J}$ satisfy the Generalized Supergravity Equations. The procedure is valid also for non-local dilaton and non-constant $\mathcal{J}$. We also show that $Div\,\Theta$ of the non-commutative structure $\Theta$ used for non-Abelian T-duality or integrable deformations does not give correct $\mathcal{J}$ for Poisson--Lie T-plurality.
\end{abstract}


\tableofcontents


\section{Introduction}

In recent years we have seen renewed interest in \natd y \cite{buscher:ssbfe,delaossa:1992vc} of sigma models. Its generalization \cite{1012.1320 sfethomp,1104.5196LCST} includes RR fields and is often used to find new supergravity solutions \cite{KlebWitt,INNSZ}. Duality transformation can be performed whenever there is an isometry of the \sm\ \bkg. However, dualization \wrt\ non-semisimple Lie groups $\G$ does not preserve conformal invariance \cite{GRV} and dual models exhibit mixed gauge and gravitational anomaly proportional to the trace of structure constants of Lie algebra $\cg$ corresponding to $\G$ \cite{aagl}.

Non-Abelian T-duality also contributes to the study of integrable deformations \cite{BorWulff1,dehato} where similar problem has been dealt with \cite{sugra2}. After publication of \cite{Wulff:2016tju} it became clear that integrable deformations should satisfy the so-called Generalized Supergavity Equations instead of usual supergravity equations. Since some deformations can be understood as non-Abelian T-duals \cite{hoatsey}, one may ask whether dual models satisfy Generalized Supergravity Equations in general.

Authors of \cite{hokico} have found non-Abelian T-duals of Bianchi
cosmologies \cite{Batakis} and have shown that dual backgrounds
indeed satisfy Generalised Supergravity Equations. In our recent
paper \cite{hlape:pltdid} we noted that beside ``full'' non-Abelian
T-dualities used in \cite{hokico} there are other \tfn s that belong
to the NATD-group investigated in \cite{LuOst}. Namely, we
considered B-shifts, $\beta$-shifts, ``factorized'' dualities and
their compositions and we have shown that \bkg s obtained by these
\tfn s satisfy Generalized Supergravity Equations as well. The above
mentioned \tfn s are best understood in context of Poisson--Lie
T-duality \cite{klise} and \pltp y \cite{unge:pltp}. Underlying
algebraic structure of these \tfn s is the \dd\ $\D$ formed by a
pair of Lie subgroups $\G$ and $\tG$.  Corresponding Lie algebras
form Manin triple $(\cd,\cg, \tcg)$.

Formulas summarized in Section \ref{basics_of_pltp} allow us to
construct \sm s on Lie groups $\G$ and $\tG$ that possess
generalized symmetries \cite{klim:proc} and satisfy Generalised
Supergravity Equations for certain dilaton field. However, beside
$\G$ and $\tG$ there might also exist other pairs of groups $\hG$
and $\bG$  that form the same \dd {} $\D$. It is then natural to ask
what \bkg s on $\hG$ or $\bG$ we get by \pltp y and whether these
\bkg s satisfy the Generalised Supergravity Equations. For
$AdS_5\times S^5$ \bkg\ this problem was addressed in \cite{saka2}.

In this paper we find Poisson--Lie T-plurals of cosmologies invariant with respect to non-simple Bianchi groups. To do so, we consider semi-Abelian \dd s $\D=(\B|\A)$, where $\B$ are three-dimensional groups of symmetries of the particular cosmology and $\A$ is three-dimensional Abelian group. Manin triples of such \dd s were classified in \cite{snohla:ddoubles}. It turns out that for each group $\B$ there are several Manin triples and many embedings of these Manin triples into $\cd$. Resulting plural \bkg s depend on many constants \cite{hlape:pltdid} and it is a tremendous task to put down complete classification. Since it is not always possible to present full results, we often restrict to particular examples of \pltp y to show important features of obtained models.

The main goal of the paper is to show that new \bkg s obtained by \pltp y satisfy Generalized Supergravity Equations. Straightforward application of dilaton \tfn\ formula \cite{unge:pltp} gives dilatons on $\hG$ and $\bG$ that may depend on coordinates of the dual group leading to puzzle discussed in \cite{snohla:puzzle}. Inspired by \cite{saka2}  we show how to deal with this issue and present formulas that allow us to calculate vector field $\mathcal{J}$ appearing in Generalized Supergravity Equations. The plural \bkg s, dilatons and $\mathcal{J}$ then satisfy these equations.

Authors of \cite{Araujo:2017jkb,Araujo:2017jap,acssy:euroj} consider Yang-Baxter deformations of $AdS_5\times S^5$ and calculate vector field $\mathcal{J}$ as $Div \Theta$ from non-commutative structure $\Theta$ in the open string picture. This works for ``full'' non-Abelian dualities, but not for ``factorized'' dualities (e.g. some \bkg s obtained in \cite{hlape:pltdid}) or more general pluralities.

Plan of the paper is following. In next Section we summarize basics of \pltp y including \tfn\ of dilaton and write down formulas for vector field $\mathcal J$. In Sections \ref{secB3}, \ref{secB5} we derive metrics, B-fields and dilatons obtained by \pltp y of  the flat background, and in Sections \ref{B_VI-1}, \ref{secB6k} we present metrics, B-fields and dilatons obtained by \pltp y of curved Bianchi cosmologies with nontrivial dilaton. It turns out that  for appropriately defined vector field $\mathcal J$ all of them satisfy Generalized Supergravity Equations.

\section{Basics of \pltp y}\label{basics_of_pltp}

In the first two subsections we recapitulate \pltp y with spectators \cite{klise,unge:pltp,hlapevoj}.

\subsection{Sigma models}

Let $\M$  be $(n+d)$-dimensional (pseudo-)Riemannian target manifold and consider sigma model on $\M$ given by Lagrangian
\begin{equation}\label{Lagrangian}
{\cal L}=\partial_- \phi^{\mu}\cf_{\mu\nu}(\phi)\partial_+ \phi^\nu,\qquad
\phi^\mu=\phi^\mu(\sigma_+,\sigma_-), \qquad
\mu=1,\ldots,n+d
\end{equation}
where tensor field $\cf=\mathcal G + \mathcal B$ on $\M$ defines
metric and torsion potential (Kalb--Ramond field) of the target
manifold. Assume that there is a $d$-dimensional Lie group $\G$ with
free action on $\M$ that leaves the tensor invariant. The action of
$\G$ is transitive on its orbits, hence we may locally consider
$\M\approx (\M/\G) \times \G = \N \times \G$, and introduce adapted
coordinates\footnote{Detailed
discussion of the process of finding adapted coordinates can be
found e.g. in \cite{hlapevoj,hlafilp:uniq}}
\begin{equation}\label{adapted}
\{s_\alpha,x^a\},\qquad \alpha=1, \ldots,n = \dim \N ,\ \ a=1,
\ldots, d = \dim \G
\end{equation}
where $s_\alpha$ label the orbits of $\G$ and are treated as
spectators, and $x^a$ are group coordinates. Dualizable sigma model
on $\N \times \G$ is given by tensor field $\mathcal{F}$ defined by
spectator-dependent $(n+d)\times (n+d)$ matrix $E(s)$  and group-dependent $\mathcal{E}(x)$ as
\begin{equation}\label{F}
\cf(s,x)=\mathcal{E}(x)\cdot E(s)\cdot \mathcal{E}^T(x), \qquad
\mathcal{E}(x)=
\left(
\begin{array}{cc}
 \unit_n & 0 \\
 0 & e(x)
\end{array}
\right)
\end{equation}
where $e(x)$ is $d\times d$ matrix of components of right-invariant Maurer--Cartan form $(dg)g^{-1}$  on $\G$.

Using non-Abelian T-duality one can find dual sigma model on
$\N\times \A$, where $\A$ is Abelian subgroup of semi-Abelian \dd\
$\D=(\G|\A)$. The necessary formulas will be given in the following
subsection as a special case of \pltp y.

In this paper the groups $\G$ will be non-semisimple Bianchi groups. Their elements will be parametrized as $g=e^{x^1 T_1}e^{x^2 T_2}e^{x^3 T_3}$ where $e^{x^2 T_2}e^{x^3 T_3}$ and $e^{x^3 T_3}$ parametrize their normal subgroups.  Bianchi cosmologies are defined on four-dimensional manifolds, hence $\dim \N=1$ and we denote the spectator $s_1$ as $t$.

\subsection{Formulas for \PL\ T-plurality with spectators}\label{pltpuvod1}

\dd \ $\D=(\G|\tG)$ is a $2d$-dimensional Lie group whose Lie
algebra $\cd$ is equipped with an ad-invariant non-degenerate
symmetric bilinear form $\langle . , . \rangle$. The Lie algebra
$\cd$ decomposes into double cross sum  $\cg \bowtie \tcg$ of
subalgebras $\cg$ and $\tcg$ \cite{majid} that are maximally
isotropic with respect to $\langle . , . \rangle$ giving rise to
Manin triple $(\cd, \cg, \tcg)$. This means that mutually dual bases
$T_a \in \cg,\ \widetilde{T}^a \in \tcg$, $a=1, \ldots, d,$ satisfy
relations
\begin{align}
\label{dual_bases}
\langle T_a, T_b \rangle & = 0, & \langle \widetilde{T}^a, \widetilde{T}^b \rangle &= 0, & \langle T_a, \widetilde{T}^b \rangle & = \delta_a^b,
\end{align}
and due to ad-invariance of the bilinear form $\langle . , . \rangle$ the algebraic structure of Manin triple is given by
\begin{equation}
\label{commutation_on_d}
[T_i,T_j]=f_{ij}^k T_k, \qquad [\widetilde T^i, \widetilde T^j]=\wt f^{ij}_k \widetilde T^k, \qquad [T_i,\widetilde T^j]=f_{ki}^j \widetilde T^k + \wt f_i^{jk} T_k.
\end{equation}

For many \dd s $\D=(\G|\tG)$ several Manin triples may exist.
Suppose that we have \sm\ on $\N\times \G$ and the \dd {} is formed
by another pair of subgroups $\hG$ and $\bG$ corresponding to Manin
triple $\hcg \bowtie \bcg$. Then we can apply the full framework of
\PL\ T-plurality \cite{klise, unge:pltp} and find backgrounds for
sigma models on $\N \times \hG$ and $\N \times \bG$. Relation
between groups $\G$,$\tG$ and $\hG$, $\bG$ is given by relation
between the Manin triples. In terms of group elements it reads
\begin{equation*}
 l=g(y)\wt h(\wt y)=\wh g(\hat x)\bar h(\bar x),\quad l\in\D, \
g\in \G,\ \wt h\in \wt\G,\ \wh g\in \wh\G,\ \bar h\in \bar\G.
\end{equation*}
The mutually dual bases $\wh T_a \in \hcg,\ \wb{T}^a \in
\hcg$, $a=1, \ldots, d,$ satisfying
\begin{align}
\label{dual_bases2}\langle \wh T_a, \wh T_b \rangle & = 0, & \langle \wb{T}^a, \wb{T}^b \rangle & = 0, & \langle \wh T_a, \wb{T}^b \rangle & =  \delta_a^b,\\
\label{commutation_on_d2}
[\wh T_i, \wh T_j]&=\hat f_{ij}^k \wh T_k, & [\wb T^i, \wb T^j]&=\wb f^{ij}_k \wb T^k, & [\wh T_i,\wb T^j]&=\hat f_{ki}^j \wb T^k + \wb f_i^{jk} \wh T_k
\end{align}
then must be related to $T_a \in \cg$ and $\widetilde{T}^a \in \tcg$ by linear transformation
\begin{equation}\label{C_mat}
\begin{pmatrix}
\wh T \\
\wb T
\end{pmatrix}
 = C \cdot
\begin{pmatrix}
T \\
\widetilde T
\end{pmatrix}
\end{equation}
where $C$ is an invertible $2d\times 2d$ matrix.
Given the structure constants $F_{ij}^k$ of $\cd=\cg \bowtie \tcg$ and $\wh F_{ij}^k$ of $\cd=\hcg \bowtie \bcg$, the matrix $C$ has to satisfy equation
\begin{equation}
\label{C_str_const}
C_a^p C_b^q F_{pq}^r = \wh F_{ab}^c C_c^r.
\end{equation}
To preserve the bilinear form $\langle . , . \rangle$ and thus \eqref{dual_bases} and \eqref{dual_bases2}, $C$ also has to satisfy
\begin{equation}
\label{C_bilin}
C_a^p C_b^q (D_0)_{pq}=(D_0)_{ab}
\end{equation}
where $(D_0)_{ab}$ are components of matrix $D_0$ that can be
written in block form as
\begin{equation}
\label{Delta} D_0=\begin{pmatrix} \mathbf{0}_d & \mathbf{1}_d \\
\mathbf{1}_d & \mathbf{0}_d \end{pmatrix}.
\end{equation}
In other words, $C$ is an element of $O(d,d)$ but, unlike the case of Abelian T-duality, not every element of $O(d,d)$ is allowed in \eqref{C_mat}.

For the following formulas it will be convenient to introduce $d \times d$ matrices $P, Q, R, S$ as
\begin{equation}\label{pqrs}
\begin{pmatrix}
T \\
\widetilde T
\end{pmatrix}
= C^{-1} \cdot
\begin{pmatrix}
\wh T \\
\wb T
\end{pmatrix} =
\begin{pmatrix}
 P & Q \\
 R & S
\end{pmatrix} \cdot
\begin{pmatrix}
\wh T \\
\wb T
\end{pmatrix}
\end{equation}
and extend these to $(n+d)\times (n+d)$ matrices
\begin{equation}
\label{pqrs2}
\nonumber
\mathcal{P} =\begin{pmatrix}\unit_n &0 \\ 0&P \end{pmatrix}, \qquad \mathcal{Q} =\begin{pmatrix}\nul_n&0 \\ 0&Q \end{pmatrix}, \qquad \mathcal{R} =\begin{pmatrix}\nul_n&0 \\ 0&R \end{pmatrix}, \qquad \mathcal{S} =\begin{pmatrix}\unit_n &0 \\ 0& S \end{pmatrix}
\end{equation}
{to accommodate the spectator fields.} It is also advantageous to
introduce block form of $E(s)$ as
\begin{equation}
\nonumber
 E(s)=
\left(
\begin{array}{cc}
 E_{\alpha\beta}(s) & E_{\alpha b}(s) \\
 E_{a\beta}(s)      & E_{ab}(s)
\end{array}
\right), \qquad
\alpha, \beta=1,\ldots,n, \quad a, b=1, \ldots, d.
\end{equation}
The sigma model on $\N \times \hG$ related to \eqref{F} via \pltp y
is given by tensor
{\begin{equation} \label{Fhat} \widehat{\cf}(s,\hat x)=
\mathcal{\widehat E}(\hat x)\cdot \widehat E(s,\hat x) \cdot
\mathcal{\widehat E}^T(\hat x), \qquad \mathcal{\widehat E}(\hat x)=
\begin{pmatrix}
\unit_n & 0 \\
 0 & \wh e(\hat x)
\end{pmatrix}
\end{equation}
where $\wh e(\hat x)$ is $d\times d$ matrix of components of
right-invariant Maurer--Cartan form $(d\hat g)\hat g^{-1}$  on
$\wh\G$,}
\begin{equation}\label{Fhat2} \wh
E(s,\hat x)=\left(\unit_{n+d}+\wh E(s) \cdot \wh{\Pi}(\hat x)\right)^{-1}\cdot
\wh E(s), 
\end{equation}
$$\widehat\Pi(\hat x)= \left(
\begin{array}{cc}
\nul_n & 0 \\
 0 & \widehat b(\hat x) \cdot \widehat a^{-1}(\hat x)
\end{array}
\right),$$
and matrices $\widehat b(\hat x)$ and $\widehat a(\hat x)$ are
submatrices of the adjoint representation
$$
ad_{{\hat g}\-1}(\wb T) = \widehat b(\hat x) \cdot \wh T + \widehat a^{-1}(\hat x)\cdot\wb T.
$$
The matrix $\wh E(s)$ is obtained from $E(s)$ in \eqref{F} by formula
\begin{equation}\label{E0hat}
\wh E(s)=(\mathcal{P}+ E(s) \cdot \mathcal{R})^{-1} \cdot
(\mathcal{Q}+E(s) \cdot \mathcal{S})
\end{equation}
so it is necessary that\footnote{For regular $\wh E (s)$ formula \eqref{Fhat2} simplifies to $\left(\wh E^{-1}(s)+ \wh{\Pi}(\hat x)\right)^{-1}$ and
$\det\, \left(\mathcal{Q}+E(s)\cdot \mathcal{S}\right)\neq 0$ is also needed.}
\begin{equation}\label{cond_plural}
\det\, \left(\mathcal P+E(s)\cdot \mathcal R \right)\neq 0.
\end{equation}
Analogously we get tensor dual to \eqref{Fhat} on $\N \times
\bar\G$.

Formulas \eqref{Fhat}--\eqref{E0hat} reduce to those for full \pltd
y if we choose $P=S=\mathbf{0}_d$ and $Q=R=\mathbf{1}_d$.
Furthermore, for a semi-Abelian \dd\ the well-known Buscher rules
for \natd y are restored. If there are no spectators the plurality
is called atomic.

\subsection{Generalized Supergravity Equations and transformation of dilaton}\label{SUGRA}

Main goal of this paper is to verify whether \bkg s obtained by \pltp ies satisfy Generalized Supergravity Equations of Motion   \cite{sugra2,Wulff:2016tju,sugra1,muck}. They can be written in different forms. We adopt convention used in \cite{hokico} so the equations read\footnote{We consider here only bosonic fields in the \bkg s.}
\begin{align}\label{betaG}
0 &= R_{\mu\nu}-\frac{1}{4}H_{\mu\rho\sigma}H_{\nu}^{\
\rho\sigma}+\nabla_{\mu}X_{\nu}+\nabla_{\nu}X_{\mu},\\ \label{betaB}
0 &=
-\frac{1}{2}\nabla^{\rho}H_{\rho\mu\nu}+X^{\rho}H_{\rho\mu\nu}+\nabla_{\mu}X_{\nu}-\nabla_{\nu}X_{\mu},\\
\label{betaPhi} 0 &=
R-\frac{1}{12}H_{\rho\sigma\tau}H^{\rho\sigma\tau}+4\nabla_{\mu}X^{\mu}-4X_{\mu}X^{\mu}
\end{align}
where
$$
H_{\rho\mu\nu}=\partial_\rho \mathcal B_{\mu\nu}+\partial_\mu  \mathcal B_{\nu\rho}+\partial_\nu  \mathcal B_{\rho\mu},
$$
$\nabla_\mu$ are covariant derivatives \wrt {} metric $\mathcal G$,
\begin{equation} \nonumber
X_{\mu}=\partial_{\mu}\Phi + \mathcal J^{\nu} \cf_{\nu\mu},
\end{equation}
$\Phi$ is the dilaton and vector field $ \mathcal J$ will be defined bellow. For vanishing $ \mathcal J$ the usual \vbe\ are recovered.

Formula for transformation of dilaton under \pltp y was given in \cite{unge:pltp} as
\begin{equation}
\wh\Phi(\hat x)= \Phi^{0}(y)-\half\ln \Big| \det \left[\left(N +
\wh\Pi(\hat x) M\right)\wh a(\hat x)\right]\Big|
\label{dualdil}
\end{equation}
where $y$ represent \coor s of $\G$,
\begin{equation}\label{phi0}
\Phi^{0}(y)=\Phi(y)+\half \ln\Big| \det \left[\left({\bf 1} + \Pi(y)
E(s)\right) a(y)\right]\Big|,
\end{equation}$\Phi(y)$ is the dilaton of the initial model and
$$ M=\mathcal{S}^T E(s)-\mathcal{Q}^T\,, \qquad
N=\mathcal{P}^T-\mathcal{R}^T E(s).
$$

Problem with the formula \eqref{dualdil} is that in general it is
non-local in the sense that $y=y(\hat x,\bar x)$ obtained by decompositions of \dd {} elements
\begin{equation}
\label{lgh}
 l=g(y)\wt h(\wt y)=\wh g(\hat x)\bar h(\bar x),\quad l\in\D, \
g\in \G,\ \wt h\in \wt\G,\ \wh g\in \wh\G,\ \bar h\in \bar\G.
\end{equation}
This is the cause of the puzzle mentioned in \cite{snohla:puzzle}, namely that
$\wh\Phi$ in formula \eqref{dualdil} for dilaton on $\hG$ depends not only on \coor s
$\hat x$ of the group $\wh\G$ but also on \coor s $\bar x$ of
$\bar{\G}$.

For coordinates $y$ depending linearly on $\hat x$ and $\bar x$
\begin{equation}
\nonumber
 y^k(\hat x,\bar x)= \hat d_m^k \hat x^m+ \bar d^{k m}\bar x_m,
\end{equation}
which holds for all cases  bellow, the problem can be solved in  the following way. We set
\begin{equation}\nonumber
\wh\Phi^{0}(y)=\Phi^{0}(\hat d_m^k \hat x^m)
\end{equation}
and vectors $\wh{\mathcal{J}}$ for \bkg s on $\N\times\wh\G$
$$\wh{\mathcal{J}}^\alpha=0,\qquad \alpha=1,\ldots, \dim \N,$$
\begin{equation}\label{kilJ}
 \mathcal{\wh J}^{\dim \N+ m} = \left(\half{{\bar f}^{ab}}{}_b -\frac{\partial{\Phi^{0}}}{\partial y^k}\bar d^{k a}\right){\wh V_a}^m, \qquad a,b,k,m=  1,
\ldots,\dim \G.
\end{equation}
For the dual model on $\N\times\bar\G$ 
\begin{equation} \nonumber
\bar\Phi^{0}(y)=\Phi^{0}(\bar d^{k m}\bar x_m),
\end{equation}
\begin{equation}\label{kilJbar}
\mathcal {\bar J}^\alpha=0,\quad  \mathcal{\bar J}^{\dim \N+ m} = \left(\half{{\hat f}_{ab}}{}^b -\frac{\partial{\Phi^{0}}}{\partial y^k}\hat d_a^k \right)\bar V^{a,m},
\end{equation}
where ${\wh V_a}$, $\bar V^{a}$ are left-invariant fields of the groups $\wh\G$, $\bar\G$ and ${\hat f}_{ba}{}^c$ and ${\bar f}^{ba}{}_c$ are structure constants of their Lie algebras. Note that in case of \pltp y linear combinations of left-invariant vector fields  need not be Killing vectors of plural \bkg s as they satisfy condition \cite{klise} of pluralizability
\begin{equation}       \label{kseq}
  (\mathfrak{L}_{\wh V_{a}}\wh \cf)_{\mu\nu} =
   \wh \cf_{\mu\kappa}\wh V^{\kappa}_{b}\bar{f}^{bc}_{a} \wh V^{\lambda}_{c}\wh \cf_{\lambda\nu}
 \end{equation}
with generally nonvanishing right-hand side. Nevertheless, it turns out that all nontrivial vectors $\mathcal {\wh J}$ and $\mathcal {\bar J}$ found below are Killing vectors of corresponding \bkg s.

{For non-Abelian T-duality, authors of \cite{Araujo:2017jkb}, \cite{Sakamoto:2017cpu} calculate components $\mathcal{ J}^\mu$ of vector $\mathcal J$ using
\begin{equation}\label{Divtheta}
(Div\,\Theta)^\mu =D_\nu \Theta^{\nu\mu}
\end{equation}
where\footnote{Note that $\Theta $ is the antisymmetric part of $\cf^{-1}(s,x)$.}
$$ \Theta^{\mu\nu}=-((\mathcal G + \mathcal B)\-1\cdot \mathcal B\cdot(\mathcal G -\mathcal B)\-1)^{\mu\nu},$$
$\mathcal G$ and $\mathcal B$ are symmetric and antisymmetric part of tensor $\cf$ and $D_\nu$ is covariant derivative \wrt
$$ G_{\mu\nu}=(\mathcal G - \mathcal B\cdot\mathcal G\-1\cdot\mathcal B)_{\mu\nu}.$$
Unfortunately, as we shall show on many examples below, this formula does not work in general for \pltp y.}
\section{\pltp y of flat \bkg}

\subsection{Bianchi $III$ cosmology}\label{secB3}

First we shall study \pltp ies that follow from the Bianchi $III$ invariance of Minkowski metric. Consider six-dimensional semi-Abelian \dd\footnote{By ${\B}_{III}$ and ${\mathfrak b}_{III}$ we denote the group Bianchi ${III}$ resp. its Lie algebra and by $\A$ and $\mathfrak a$ we denote three dimensional
Abelian group resp. its Lie algebra. Similar notation
will be used in the following sections.}  $\D=({\B}_{III} |\A)$ whose Lie
algebra $\mathfrak{d}=\mathfrak b_{III}\bowtie \mathfrak a$ is spanned by
basis $(T_1,T_2,T_3,\wwt T^1,\wwt T^2,\wwt T^3)$ {with non-vanishing commutation} relations
\begin{equation}
\label{MT31}[ T_1, T_3]=- T_3, \qquad [
T_1, \wwt T^3]= \wwt T^3, \qquad [ T_3, \wwt T^3]= -\wwt T^1.
\end{equation}
The group ${\B}_{III}$ is not semisimple and trace of its structure
constants does not vanish. Lie algebra $\cd$ of this \dd\ admits several other decompositions into Manin triples found in \cite{snohla:ddoubles}.

In adapted coordinates the Bianchi $III$ flat cosmology is given by metric\footnote{For typographic reasons from now on we shall use subscripts to label group coordinates.}
\begin{equation} \label{F3_orig}
\cf (t,y_1) = \left(
\begin{array}{cccc}
 -1 & 0 & 0 & 0 \\
 0 & t^2 & 0 & 0 \\
 0 & 0 & 1 & 0 \\
 0 & 0 & 0 & e^{- 2 y_1} t^2 \\
\end{array}
\right).
\end{equation}
The corresponding matrix $E(s)$ can be found by setting $y_1=0$ in $\cf$.
The {\bkg\ }is invariant \wrt\ the action of Bianchi $III$ group
generated by left-invariant vector fields
\begin{equation}\label{linver3}
V_1=\partial_{y_1}+y_3\,\partial_{y_3}, \qquad
V_2=\partial_{y_2}, \qquad V_3=\partial_{y_3}.
\end{equation}
The metric is flat and there is no torsion so the equations \eqref{betaG}--\eqref{betaPhi} are satisfied if we choose dilaton $\Phi(y)=0$ and vector $\mathcal{J}=0$. This \bkg\ was mentioned in \cite{GR} where the authors noted that its non-Abelian dual does not satisfy standard \vbe.

The \eqn\ \eqref{phi0} implies $\Phi^{(0)}(y)=\frac{y_1}{2} $ and formula \eqref{dualdil} for new dilatons then simplifies to
\begin{equation}
\wh\Phi(\hat x, \bar x)=\frac{y_1(\hat x, \bar x)}{2}-\half\ln \Big| \det \left[\left(N + \wh\Pi(\hat x) M\right)\wh a(\hat x)\right]\Big|.
\label{dualdil31}
\end{equation}
Hence, for every $\hat{\mathfrak g}\bowtie\bar{ \mathfrak g}$ it is necessary to express $y_1$ in terms of $\hat x$ and $\bar x$ from \eqref{lgh}.

\subsubsection{Transformation of $\mathfrak b_{III}\bowtie \mathfrak a$ to $\mathfrak b_{III}\bowtie \mathfrak b_{II}$ and to its dual}

As was found in \cite{snohla:ddoubles}, the algebra of the \dd\ $\D=({\B}_{III} |\A)$ can be decomposed into Manin triple $\mathfrak{d}=\mathfrak b_{III}\bowtie \mathfrak b_{II}$ spanned by basis $(\wh T_1,\wh T_2,\wh T_3,\wb T^1,\wb T^2,\wb T^3)$ with non-trivial algebraic relations
\begin{align}
\label{com3132} [ \wh T_1, \wh T_2] & = -\wh T_2, & [ \wh T_1, \wb T^2] & = (\wh T_3+\wb T^2), & [ \wh T_1, \wb T^3] & = -\wh T_2, \\  & & [ \wh T_2, \wb T^2] & = -\wb T^1, & [ \wb T^2, \wb T^3] & = \wb T^1. \nonumber
\end{align}
Left-invariant fields of $\wh\G={\B}_{III}$ and $\bar\G={\B}_{II}$ are
\begin{align}\nonumber 
\wh V_1 & =\partial_{\hat x_1}+\hat x_2\,\partial_{\hat x_2}, & \wh
V_2 & =\partial_{\hat x_2}, & \wh V_3 & =\partial_{\hat x_3}, \\
\nonumber 
\bar V_1 & =\partial_{\bar x_1}, & \bar V_2 & =-\bar x_3\,\partial_{\bar x_1}+\,\partial_{\bar x_2}, & \bar V_3 & =\partial_{\bar x_3}.
\end{align}

There is rather large number of solutions of equations \eqref{C_str_const} and \eqref{C_bilin} giving linear mappings \eqref{C_mat} between {generators} of $\mathfrak b_{III}\bowtie \mathfrak a$ and $\mathfrak b_{III}\bowtie \mathfrak b_{II}$ (cf. the table of \PL\ identities of $\mathfrak b_{III}\bowtie \mathfrak a$ in \cite{hlape:pltdid}). However, up to a change of coordinates and gauge shift of the $\wh{\mathcal{B}}$ field, all the resulting plural \bkg s $\wh\cf$ on $\hG$ are equivalent. Since the parameters coming from mappings $C$ can be eliminated in plural backgrounds by transformations of coordinates or gauge shifts, it is sufficient to consider matrix
\begin{equation} \label{C31328}
C_1=\left(
\begin{array}{cccccc}
 1 & 0 & 0 & 0 & 0 & 0 \\
 0 & 0 & 1 & 0 & 0 & 0 \\
 0 & 1 & 0 & 0 & 0 & 0 \\
 0 & 0 & 0 & 1 & 0 & 0 \\
 0 & -1 & 0 & 0 & 0 & 1 \\
 0 & 0 & 1 & 0 & 1 & 0 \\
\end{array}
\right),
\end{equation}
which gives rise to tensor $\wh{\cal F}$ on $\wh\G$ calculated from
\eqref{Fhat},\eqref{Fhat2} as
\begin{equation} \nonumber
\wh{\cal F}(t,\hat x_1)
=\left(
\begin{array}{cccc}
 -1 & 0 & 0 & 0 \\
 0 & t^2 & 0 & 0 \\
 0 & 0 & \frac{t^2}{e^{2 \hat x_1}+t^2} & -\frac{t^2}{e^{2 \hat x_1} +t^2} \\
 0 & 0 & \frac{t^2}{e^{2 \hat x_1} +t^2} & \frac{e^{2 \hat x_1} }{e^{2 \hat x_1} +t^2} \\
\end{array}
\right).
\end{equation}
The metric is not flat but its scalar curvature vanishes. It is possible to find coordinate transformation that brings the metric to the Brinkmann form of a plane parallel wave
$$
ds^2=\frac{\left(2 u^2 -1\right) z_3^2-3 z_4^2}{\left(1+u^2\right)^2}du^2+2du\,dv+d z_3^2+d z_4^2.
$$
The \bkg\ is accompanied by a non-trivial torsion
$$
\wh H = \frac{2}{1+u^2}du\wedge dz_3 \wedge dz_4.
$$
This \bkg\ was repeatedly found in \cite{hp,php,hpp} as non-Abelian T-dual of flat metric.

Let us now focus on the dilaton. Change of Manin triples given by mapping \eqref{C31328} induces \tfn {} of \coor s of the \dd. {From \eqref{lgh} we find that}
\begin{align*}
y_1 &= \hat x_1, & y_2 &= \hat x_3-\bar x_2, & y_3 &= \hat x_2+\bar x_3,\\
\wt y_1 &= \bar x_1-\bar x_2\bar x_3, & \wt y_2 &= \bar x_3, & \wt y_3 &= \bar x_2.
\end{align*}
As  $\Phi^{(0)}(y)=\frac{y_1}{2}$, we need only $y_1$ and that gives us
$\Phi^{(0)} =\frac{\hat x_1}{2}$. From the formula \eqref{dualdil} we obtain transformed dilaton
\begin{equation} \nonumber
\wh\Phi(t,\hat x_1)=-\frac{1}{2} \ln \left(t^2 e^{-2 {\hat x_1}}+1\right).
\end{equation}
Vector $\mathcal {\wh J}$ given by both \eqref{kilJ} and \eqref{Divtheta} vanishes,  Generalized Supergravity Equations become standard \vbe, and one can check that they are satisfied.

For sigma model on $\bG$ different $C$ matrices give different \bkg s. For instance, using
\begin{equation} \nonumber
C_2=D_0\cdot C_1=\left(
\begin{array}{cccccc}
 0 & 0 & 0 & 1 & 0 & 0 \\
 0 & -1 & 0 & 0 & 0 & 1 \\
 0 & 0 & 1 & 0 & 1 & 0 \\
 1 & 0 & 0 & 0 & 0 & 0 \\
 0 & 0 & 1 & 0 & 0 & 0 \\
 0 & 1 & 0 & 0 & 0 & 0 \\
\end{array}
\right)
\end{equation}
we obtain tensor
\begin{equation} \nonumber
\wb{\cal F}(t,\bar x_2)
=\left(
\begin{array}{cccc}
 -1 & 0 & 0 & 0 \\
 0 & \frac{t^2}{t^4+\bar x_2^2} & -\frac{\bar x_2}{t^4+\bar x_2^2} & 0 \\
 0 & \frac{\bar x_2}{t^4+\bar x_2^2} & \frac{t^2}{t^4+\bar x_2^2} & 1 \\
 0 & 0 & -1 & 1 \\
\end{array}
\right)
\end{equation}
with non-trivial curvature and torsion. To find the dilaton and vector $\wb{\mathcal J}$ we express $y_1$ from \eqref{lgh} as $y_1=\hat x_1$. Dilaton
\begin{equation} \nonumber
\wb\Phi(t,\bar x_2)=-\frac{1}{2} \ln \left(t^4+\bar x_2^2\right)
\end{equation}
found by \pltp y {generated by $C_2 = D_0\cdot C_1$} transforming $\mathfrak b_{III}\bowtie \mathfrak a$ to $\mathfrak b_{II}\bowtie \mathfrak b_{III}$ satisfies Generalized Supergravity Equations {if the vector}
$$
\wb{\mathcal J}= (0,-1,0,0)
$$
is calculated via \eqref{kilJbar}. Formula \eqref{Divtheta} fails to provide correct $\bar{\mathcal{J}}$ as
$$
D_{\nu}\Theta^{\nu \mu}= \left(0,-\frac{1}{1+t^2},0,0\right).
$$

Using
\begin{equation} \nonumber
C_3=\left(
\begin{array}{cccccc}
 0 & 0 & 0 & -1 & 0 & 0 \\
 0 & 0 & 1 & 0 & -1 & 0 \\
 0 & 1 & 0 & 0 & 0 & 1 \\
 -1 & 0 & 0 & 0 & 0 & 0 \\
 0 & 0 & 0 & 0 & 0 & 1 \\
 0 & 0 & 0 & 0 & 1 & 0 \\
\end{array}
\right)
\end{equation}
we get torsionless background
\begin{equation} \nonumber
\wb{\cal F}(t,\bar x_2)
=\left(
\begin{array}{cccc}
 -1 & 0 & 0 & 0 \\
 0 & \frac{1}{\left(\bar x_2^2+1\right) t^2} & -\frac{\bar x_2}{\bar x_2^2+1} & 0 \\
 0 & \frac{\bar x_2}{\bar x_2^2+1} & \frac{t^2}{\bar x_2^2+1} & 1 \\
 0 & 0 & -1 & 1 \\
\end{array}
\right).
\end{equation}
From \eqref{lgh} we find that $y_1=-\hat x_1$ and formulas \eqref{dualdil} and \eqref{kilJbar} give
\begin{equation} \nonumber
\wb\Phi(t,\bar x_2)=-\frac{1}{2} \ln \left(t^2(1+\bar x_2^2)\right), \qquad \wb{\mathcal J}= (0,0,0,0).
\end{equation}
The standard \vbe\ are thus satisfied. Since $D_{\nu}\Theta^{\nu \mu}= \left(0,-\frac{t^2}{1+t^2},0,0\right)$, we see that formula \eqref{Divtheta} is again not applicable.

\subsubsection{Transformation of $\mathfrak b_{III}\bowtie \mathfrak a$ to $\mathfrak b_{III}\bowtie \mathfrak b_{IIIiii}$ and to its dual}

Lie algebra $\mathfrak{d}=\mathfrak b_{III}\bowtie \mathfrak b_{IIIiii}$
is spanned by $(\wh T_1,\wh T_2,\wh T_3,\wb T^1,\wb T^2,\wb T^3)$
with non-trivial algebraic relations
$$
[ \wh T_1, \wh T_2]= -4\,\wh T_2,\qquad [ \wh T_1, \wb T^1]= 2\,\wh T_2, \qquad [ \wh T_1, \wb T^2]= (-2\,\wh T_1+4\,\wb T^2),
$$
\begin{equation} \nonumber
[ \wh T_2, \wb T^2]= - 4\, \wb T^1, \qquad [ \wb T^1, \wb T^2]=2\, \wb T^1.
\end{equation}
The left-invariant fields of $\wh\G$ and $\bar\G$ are
\begin{align*}
\wh V_1 & =\partial_{\hat x_1}+4 \hat x_2\,\partial_{\hat x_2}, & \wh
V_2 & =\partial_{\hat x_2}, & \wh V_3 & =\partial_{\hat x_3}, \\
\bar V_1 & =\e^{-2 \bar x_2}\partial_{\bar x_1}, & \bar
V_2 & =\partial_{\bar x_2}, & \bar V_3 & =\partial_{\bar x_3}.
\end{align*}
There is again rather large number of linear mappings \eqref{C_mat} between  $\mathfrak b_{III}\bowtie \mathfrak a$ and $\mathfrak b_{III}\bowtie \mathfrak b_{IIIiii}$ depending on many parameters. The \bkg s obtained by plurality are quite complicated, hence, we shall restrict to two simple ones.

First of them is given by matrix 
\begin{equation} \nonumber 
C_1=\left(
\begin{array}{cccccc}
 -4 & 0 & 0 & 0 & 0 & 0 \\
 0 & 0 & 0 & 0 & 0 & 1 \\
 0 & 1 & 0 & 0 & 0 & 0 \\
 0 & 0 & 0 & -\frac{1}{4} & 0 & -\frac{1}{2} \\
 -2 & 0 & 1 & 0 & 0 & 0 \\
 0 & 0 & 0 & 0 & 1 & 0 \\
\end{array}
\right)
\end{equation}
for which tensor $\wh \cf$ has the form
\begin{equation} \nonumber
\wh{\cal F}(t,\hat x_1)
=\left(
\begin{array}{cccc}
 -1 & 0 & 0 & 0 \\
 0 & \frac{16 e^{8 {\hat x_1}} t^2}{4+e^{8 {\hat x_1}}} &
   -\frac{8}{4+e^{8 {\hat x_1}}} & 0 \\
 0 & \frac{8}{4+e^{8 {\hat x_1}}} & \frac{1}{\left(4+e^{8
   {\hat x_1}}\right) t^2} & 0 \\
 0 & 0 & 0 & 1 \\
\end{array}
\right).
\end{equation}
The \bkg\ is curved and torsionless. From the change of decompositions \eqref{lgh} we express $y_1$ as
$$
y_1=-4 \hat x_1-2\bar x_2 .
$$
Dilaton
\begin{equation} \nonumber
\wh\Phi(t,\hat x_1)=-\frac{1}{2} \ln \left(\frac{t^2}{4}(4+\, e^{8{\hat x_1}})\right)
\end{equation}
satisfies Generalized Supergravity Equations with $\wh{\mathcal J}=  (0,0,0,0)$ but $Div\,\Theta= (0,0,-2,0)$.

Torsionless background
\begin{equation} \nonumber
\wb{\cal F}(t,\bar x_1,\bar x_2)
=\left(
\begin{array}{cccc}
 -1 & 0 & 0 & 0 \\
 0 & \frac{e^{4 {\bar x_2}}}{4 \left(5-2 e^{2 {\bar x_2}}+e^{4
   {\bar x_2}}\right) t^2} & \frac{5 e^{2 {\bar x_2}} t^2+e^{4 {\bar x_2}}
   \left({\bar x_1}-t^2\right)}{2 \left(5-2 e^{2 {\bar x_2}}+e^{4
   {\bar x_2}}\right) t^2} & 0 \\
 0 & \frac{e^{4 {\bar x_2}} \left(t^2+{\bar x_1}\right)-5 e^{2 {\bar x_2}}
   t^2}{2 \left(5-2 e^{2 {\bar x_2}}+e^{4 {\bar x_2}}\right) t^2} &
   \frac{e^{4 {\bar x_2}} \left(4 t^4+{\bar x_1}^2\right)}{\left(5-2
   e^{2 {\bar x_2}}+e^{4 {\bar x_2}}\right) t^2} & 0 \\
 0 & 0 & 0 & 1 \\
\end{array}
\right)
\end{equation}
is obtained by \pltp y given by $D_0\cdot C_1$. Transforming $\mathfrak b_{III}\bowtie \mathfrak a$ to $\mathfrak b_{IIIiii}\bowtie \mathfrak b_{III}$ we get
dilaton
\begin{equation} \nonumber
\wb\Phi(t,\bar x_2)=-\frac{1}{2} \ln \left(t^2 \left(-2 e^{2 {\bar x_2}}+e^{4 {\bar x_2}}+5\right)\right)
\end{equation}
that satisfies \vbe, i.e. Generalized Supergravity Equations with $\mathcal{\bar J}= (0,0,0,0)$. That disagrees with formula \eqref{Divtheta} giving $Div\,\Theta = (0,-4 e^{-2 {\bar x_2}},0,0)$.

Another \pltp y is given by the matrix
\begin{equation} \nonumber 
C_2=\left(
\begin{array}{cccccc}
 4 & 0 & 0 & 0 & 0 & 0 \\
 0 & 0 & 2 & 0 & 0 & 0 \\
 0 & 1 & 0 & 0 & 0 & 0 \\
 0 & 0 & -1 & \frac{1}{4} & 0 & 0 \\
 2 & 0 & 0 & 0 & 0 & \frac{1}{2} \\
 0 & 0 & 0 & 0 & 1 & 0 \\
\end{array}
\right)
\end{equation}
that leads to \bkg
\begin{equation} \nonumber
\wh{\cal F}(t,\hat x_1)
=\left(
\begin{array}{cccc}
 -1 & 0 & 0 & 0 \\
 0 & \frac{16 e^{8 {\hat x_1}} t^2}{16 t^4+e^{8 {\hat x_1}}} &
   -\frac{32 t^4}{16 t^4+e^{8 {\hat x_1}}} & 0 \\
 0 & \frac{32 t^4}{16 t^4+e^{8 {\hat x_1}}} & \frac{4 t^2}{16
   t^4+e^{8 {\hat x_1}}} & 0 \\
 0 & 0 & 0 & 1 \\
\end{array}
\right)
\end{equation}
with nontrivial torsion. For $C_2$ we find that
$$
y_1=4\hat x_1+2\bar x_2 ,\quad \Phi^{(0)} =2{\hat x_1},
$$
and the \bkg\ supported by dilaton
\begin{equation} \nonumber
\wh\Phi(t,\hat x_1)=-\frac{1}{2} \ln \left(2\, t^4 e^{-8{\hat x_1}}+\frac{1}{8}\right)
\end{equation}
satisfies Generalized Supergravity Equations with $\wh{\mathcal J}= (0,0,-2,0)$ obtained by both \eqref{kilJ} and \eqref{Divtheta}.
Background obtained by the \pltp y given by $D_0\cdot C_2$  transforming
$\mathfrak b_{III}\bowtie \mathfrak a$ to $\mathfrak b_{IIIiii}\bowtie \mathfrak b_{III}$ is
\begin{align}\label{hatF3iii31}
&\wb{\cal F}(t,\bar x_1, \bar x_2)
=\\ \nonumber & \left(
\begin{array}{cccc}
 -1 & 0 & 0 & 0 \\
 0 & \frac{e^{4 {\bar x_2}} t^2}{16 t^4-2 e^{2 {\bar x_2}}+e^{4
   {\bar x_2}}+1} & \frac{e^{2 {\bar x_2}} \left(16 t^4+e^{2 {\bar x_2}}
   \left(4 t^2 {\bar x_1}-1\right)+1\right)}{2 \left(16 t^4-2
   e^{2 {\bar x_2}}+e^{4 {\bar x_2}}+1\right)} & 0 \\
 0 & \frac{e^{2 {\bar x_2}} \left(-16 t^4+e^{2 {\bar x_2}} \left(4
   {\bar x_1} t^2+1\right)-1\right)}{2 \left(16 t^4-2 e^{2
   {\bar x_2}}+e^{4 {\bar x_2}}+1\right)} & \frac{4 e^{4 {\bar x_2}} t^2
   \left(\bar x_1^2+1\right)}{16 t^4-2 e^{2 {\bar x_2}}+e^{4
   {\bar x_2}}+1} & 0 \\
 0 & 0 & 0 & 1 \\
\end{array}
\right).
\end{align}
Dilaton
\begin{equation} \nonumber
 \wb\Phi(t,\bar x_2)=-\frac{1}{2} \ln
\left(\frac{1}{2} e^{-4 {\bar x_2}} \left(16 t^4-2 e^{2
   {\bar x_2}}+e^{4 {\bar x_2}}+1\right)\right)
\end{equation}
calculated via formula \eqref{dualdil31} and \bkg {} \eqref{hatF3iii31}
satisfy Generalized Supergravity Equations with non-constant vector
\begin{equation} \nonumber 
\wb{\mathcal J} = (0,-4 e^{-2{\bar x_2}},0,0)
\end{equation} obtained from both \eqref{kilJ} and \eqref{Divtheta}. Although it is not obvious, $\wb{\mathcal{J}}$ is a Killing vector of \eqref{hatF3iii31}. This example shows that components of left-invariant fields need to appear in formulas \eqref{kilJ}, \eqref{kilJbar}. Papers \cite{hokico}, \cite{hlape:pltdid} dealt only with semi-Abelian \dd s where $\bar V^{a,m}=\delta^{a,m}$, vectors $\wb{\mathcal J}$ were constant and proportional to the trace of structure constants so the full form of \eqref{kilJbar} was not necessary.

\subsection{Bianchi $V$ cosmology}\label{secB5}

Next we shall study \pltp ies that follow from the well-known Bianchi $V$ invariance of the flat metric that in adapted \coor s reads
\begin{equation} \label{F5_orig}
\cf (t,y_1) = \left(
\begin{array}{cccc}
 -1 & 0 & 0 & 0 \\
 0 & t^2 & 0 & 0 \\
 0 & 0 & e^{2 y_1} t^2 & 0 \\
 0 & 0 & 0 & e^{2 y_1} t^2 \\
\end{array}
\right).
\end{equation}
Authors of \cite{GRV} first noticed that non-Abelian dual of this \bkg\ is not conformal. Further study of the emerging gravitational-gauge anomaly was carried out in \cite{EGRSV}.

{Metric \eqref{F5_orig} is invariant \wrt\ the action of Bianchi $V$ group and can be obtained by formula \eqref{F} if we consider six-dimensional semi-Abelian \dd\ $\D=({\B}_V |\A)$ and Manin triple
$\mathfrak{d}=\mathfrak b_V\bowtie \mathfrak a$ spanned by basis $(T_1,T_2,T_3,\wwt T^1,\wwt T^2,\wwt T^3)$ with non-trivial \comrel s}
\begin{align}
\label{MT51}
[ T_1, T_2] & = T_2, & [ T_1, T_3] & = T_3, & [T_1, \wwt T^2] & = -\wwt T^2, \\ \nonumber
[ T_1, \wwt T^3] & = -\wwt T^3, & [T_2, \wwt T^2] & = \wwt T^1, & [ T_3, \wwt T^3] & = \wwt T^1.
\end{align}
The group ${\B}_V$ is not semisimple and trace of its structure
constants does not vanish. The algebra $\cd$ allows several other decompositions into Manin triples $\hat{\mathfrak g}\bowtie\bar{ \mathfrak g}$.

Corresponding dilaton can be again chosen $\Phi = 0$ and transformation formulas \eqref{dualdil},\eqref{phi0}  now give
\begin{equation} \nonumber
\wh\Phi(\hat x, \bar x)=-y_1(\hat x, \bar x)-\half\ln \Big| \det \left[\left(N + \wh\Pi(\hat x) M\right)\wh a(\hat x)\right]\Big|
\label{dualdil51}
\end{equation}
so it is again necessary to find how $y_1$ depends on $\hat x,\bar x$ using \eqref{lgh}.

\subsubsection{Transformation of $\mathfrak b_{V}\bowtie \mathfrak a$ to $\mathfrak b_{VI_{-1}}\bowtie \mathfrak a$ and to its dual}

Lie algebra $\mathfrak{d}=\mathfrak b_{VI_{-1}}\bowtie \mathfrak a$
is spanned by $(\wh T_1,\wh T_2,\wh T_3,\wb T^1,\wb T^2,\wb T^3)$
with algebraic relations
\begin{align}
\label{com611}
[\wh T_1, \wh T_2] & = -\wh T_2, & [ \wh T_1, \wh T_3] & = \wh T_3, & [ \wh T_1, \wb T^2] & = \wb T^2,\\ \nonumber
[ \wh T_1, \wb T^3] & = -\wb T^3, & [\wh T_2, \wb T^2] & = -\wb T^1, &[ \wh T_3, \wb T^3] & = \wb T^1.
\end{align}
Left-invariant vector fields of $\wh\G$ and $\bar\G$ are
\begin{align*}
\wh V_1 & =\partial_{\hat x_1}+\hat x_2\,\partial_{\hat x_2}-\hat x_3\,\partial_{\hat x_3}, & \wh V_2 & =\partial_{\hat x_2}, & \wh V_3 & =\partial_{\hat x_3},\\
\bar V_1 & =\partial_{\bar x_1}, & \bar V_2 & =\partial_{\bar x_2}, & \bar V_3 & =\partial_{\bar x_3}.
\end{align*}

Mappings that transform commutation relations of Manin triple
$\mathfrak b_{V}\bowtie \mathfrak a$ to those of $\mathfrak b_{VI_{-1}}\bowtie \mathfrak a$ are in general given by matrices
\begin{equation} \label{C6151}
C_1=\left(
\begin{array}{cccccc}
 1 & c_{12} & c_{13} & -c_{12} c_{15}-c_{13} c_{16} & c_{15} & c_{16} \\
 0 & 0 & 0 & -c_{12} c_{25}-c_{13} c_{26} & c_{25} & c_{26} \\
 0 & \frac{c_{26}}{c_{26} c_{65}-c_{25} c_{66}} & \frac{c_{25}}{c_{25} c_{66}-c_{26}
   c_{65}} & \frac{c_{16} c_{25}-c_{15} c_{26}}{c_{26} c_{65}-c_{25} c_{66}} & 0 & 0 \\
 0 & 0 & 0 & 1 & 0 & 0 \\
 0 & \frac{c_{66}}{c_{25} c_{66}-c_{26} c_{65}} & \frac{1}{c_{26}-\frac{c_{25}
   c_{66}}{c_{65}}} & \frac{c_{16} c_{65}-c_{15} c_{66}}{c_{25} c_{66}-c_{26} c_{65}} &
   0 & 0 \\
 0 & 0 & 0 & -c_{12} c_{65}-c_{13} c_{66} & c_{65} & c_{66} \\
\end{array}
\right)
\end{equation}
and
\begin{equation} \label{C6151b}
C_2=\left(
\begin{array}{cccccc}
 -1 & c_{12} & c_{13} & c_{12} c_{15}+c_{13} c_{16} & c_{15} & c_{16} \\
 0 & \frac{c_{36}}{c_{36} c_{55}-c_{35} c_{56}} & \frac{c_{35}}{c_{35} c_{56}-c_{36}
   c_{55}} & \frac{c_{16} c_{35}-c_{15} c_{36}}{c_{35} c_{56}-c_{36} c_{55}} & 0 & 0 \\
 0 & 0 & 0 & c_{12} c_{35}+c_{13} c_{36} & c_{35} & c_{36} \\
 0 & 0 & 0 & -1 & 0 & 0 \\
 0 & 0 & 0 & c_{12} c_{55}+c_{13} c_{56} & c_{55} & c_{56} \\
 0 & \frac{c_{56}}{c_{35} c_{56}-c_{36} c_{55}} & \frac{1}{c_{36}-\frac{c_{35}
   c_{56}}{c_{55}}} & \frac{c_{16} c_{55}-c_{15} c_{56}}{c_{36} c_{55}-c_{35} c_{56}} &
   0 & 0 \\
\end{array}
\right).
\end{equation}
{Although \bkg s calculated from $C_1$ and $C_2$ seem rather complicated at first sight, dependence on constants $c_{ij}$ can be eliminated in the resulting metrics and $c_{ij}$ appear only in $\wh{\mathcal{B}}$. Moreover, the torsion vanishes in both cases, hence, up to coordinate or gauge transformations, tensors $\wh{\cf}$ are equivalent to
\begin{equation}\label{mtz61fd3}
  \wh\cf(t,\hat x_1) =  \left(
\begin{array}{cccc}
 -1 & 0 & 0 & 0 \\
 0 & t^2 & 0 & 0 \\
 0 & 0 & e^{-2 \hat x_1} t^2 & 0 \\
 0 & 0 & 0 & \frac{e^{2 \hat x_1}}{t^2} \\
\end{array}
\right).
\end{equation}
Such \bkg\ can be obtained directly if we choose $c_{12} =c_{13}=c_{15} =c_{16} =c_{35} =c_{56}=0$ and $c_{36} =c_{55}=1$ in $C_2$, in which case we perform ``factorized'' duality
\begin{equation}\label{FD4}
C_F= \left(
\begin{array}{cccccc}
 -1 & 0 & 0 & 0 & 0 & 0 \\
 0 & 1 & 0 & 0 & 0 & 0 \\
 0 & 0 & 0 & 0 & 0 & 1 \\
 0 & 0 & 0 & -1 & 0 & 0 \\
 0 & 0 & 0 & 0 & 1 & 0 \\
 0 & 0 & 1 & 0 & 0 & 0 \\
\end{array}
\right),
\end{equation}
i.e. Buscher duality\footnote{Duality is accompanied by a change of sign in $\hat{x}_1$ that is necessary to get \eqref{com611}.} in coordinate $y_3$.}

Tensor \eqref{mtz61fd3} together with dilaton
$$
\wh\Phi(t,\hat x_1)= -\ln\, t+\hat x_1
$$
found from \eqref{dualdil51} and $y_1 = - \hat x_1$ obtained using \eqref{lgh} satisfy standard \vbe {}.

Let us note that metric \eqref{mtz61fd3} can be brought to the Brinkmann form of a plane-parallel wave
$$
ds^2=\frac{2 z_3^2}{u^2}du^2+2du\,dv+d z_3^2+d z_4^2
$$
found earlier in \cite{hp,php,hpp} as non-Abelian T-dual of flat metric.

Backgrounds $\wb\cf$ on $\bG$ found from mappings $D_0\cdot C_1$ and $D_0\cdot C_2$ can be simplified substantially since after suitable coordinate transformation they differ from \bkg\
\begin{equation} \nonumber 
\wb\cf(t,\bar x_2,\bar x_3)=\left(
\begin{array}{cccc}
 -1 & 0 & 0 & 0 \\
 0 & \frac{t^2}{\left(\bar x_3^2+1\right) t^4+\bar x_2^2} & -\frac{\bar x_2}{\left(\bar x_3^2+1\right) t^4+\bar x_2^2} & \frac{\bar x_3 t^4}{\left(\bar x_3^2+1\right) t^4+\bar x_2^2} \\
 0 & \frac{\bar x_2}{\left(\bar x_3^2+1\right) t^4+\bar x_2^2} & \frac{\left(\bar x_3^2+1\right) t^2}{\left(\bar x_3^2+1\right) t^4+\bar x_2^2} & \frac{\bar x_2 \bar x_3 t^2}{\left(\bar x_3^2+1\right) t^4+\bar x_2^2} \\
 0 & -\frac{\bar x_3 t^4}{\left(\bar x_3^2+1\right) t^4+\bar x_2^2} & \frac{\bar x_2 \bar x_3 t^2}{\left(\bar x_3^2+1\right) t^4+\bar x_2^2} & \frac{t^2 \left(t^4+\bar x_2^2\right)}{\left(\bar x_3^2+1\right) t^4+\bar x_2^2} \\
\end{array}
\right)
\end{equation}
obtained using $D_0 \cdot C_F$ only by a constant shift of the $\bar{\mathcal{B}}$ field.
Since $y_1=-\hat x_1$, we find dilaton and vector $\bJ$ as
$$
\bar \Phi(t, \bar x_2, \bar x_3)= -\frac{1}{2} \ln \left({t^4 \left(\bar x_3^2+1\right)+\bar x_2^2}\right), \qquad \bJ =(0,-1,0,0)
$$
and the Generalized Supergravity Equations \eqref{betaG}-\eqref{betaPhi}  are satisfied. On the other hand, we have $Div \Theta=(0,0,0,0)\neq \bJ $.

\subsubsection{Transformation of $\mathfrak b_{V}\bowtie \mathfrak a$ to $\mathfrak b_{V}\bowtie \mathfrak b_{II}$ and to its dual}

Lie algebra $\mathfrak{d}=\mathfrak b_{V}\bowtie \mathfrak b_{II}$
is spanned by $(\wh T_1,\wh T_2,\wh T_3,\wb T^1,\wb T^2,\wb T^3)$
with {algebraic relations
\begin{align}
\label{com52}
 [ \wh T_1, \wh T_2] & = \wh T_2, & [ \wh T_1, \wh T_3] & = \wh T_3, & [ \wh T_1, \wb T^2] & = \wh T_3 - \wb T^2,\\ \nonumber
 [ \wh T_1, \wb T^3] & = -\wh T_2 - \wb T^3, & [ \wh T_2, \wb T^2] & = \wb T^1, & [ \wh T_3, \wb T^3] & = \wb T^1, & [ \wb T^2, \wb T^3] & = \wb T^1.
\end{align}}
Left-invariant vector fields of $\wh\G$ and $\bar\G$ are
\begin{align*}
\wh V_1 & =\partial_{\hat x_1}- \hat x_2\,\partial_{\hat x_2}-\hat x_3\,\partial_{\hat x_3}, & \wh V_2 & =\partial_{\hat x_2}, & \wh V_3 & =\partial_{\hat  x_3},\\
\bar V_1 & =\partial_{\bar x_1}, & \bar V_2 & =-\bar x_3\partial_{\bar x_1}+\,\partial_{\bar x_2}, & \bar V_3 & =\partial_{\bar x_3}.
\end{align*}

Mappings $C$ that transform the algebraic relations of Manin triple
$\mathfrak b_{V}\bowtie \mathfrak a$ to $\mathfrak
b_{V}\bowtie \mathfrak b_{II}$ are given by matrices
\begin{equation} \nonumber 
C_1=\left(
\begin{array}{cccccc}
 1 & c_{12} & c_{13} & -c_{12} c_{15}-c_{13} c_{16} &
   c_{15} & c_{16} \\
 0 & c_{22} & c_{23} & -c_{15} c_{22}-c_{16} c_{23} & 0 &
   0 \\
 0 & c_{32} & c_{33} & -c_{15} c_{32}-c_{16} c_{33} & 0 &
   0 \\
 0 & 0 & 0 & 1 & 0 & 0 \\
 0 & \frac{c_{32}}{2} & \frac{c_{33}}{2} & \Gamma_{54} & \frac{c_{33}}{c_{22} c_{33}-c_{23} c_{32}} &
   \frac{c_{32}}{c_{23} c_{32}-c_{22} c_{33}} \\
 0 & -\frac{c_{22}}{2} & -\frac{c_{23}}{2} & \Gamma_{64}&
   \frac{c_{23}}{c_{23} c_{32}-c_{22} c_{33}} &
   \frac{c_{22}}{c_{22} c_{33}-c_{23} c_{32}} \\
\end{array}
\right)
\end{equation}
where $$ \Gamma_{54}=\frac{2 c_{13}
   c_{32}+c_{15} \left(c_{23} c_{32}-c_{22} c_{33}\right)
   c_{32}+c_{33} \left(c_{16} \left(c_{23} c_{32}-c_{22}
   c_{33}\right)-2 c_{12}\right)}{2 c_{22} c_{33}-2 c_{23}
   c_{32}},$$
   $$ \Gamma_{64}=\frac{-2
   c_{13} c_{22}+2 c_{12} c_{23}+\left(c_{15}
   c_{22}+c_{16} c_{23}\right) \left(c_{22} c_{33}-c_{23}
   c_{32}\right)}{2 c_{22} c_{33}-2 c_{23} c_{32}},$$
and
\begin{equation} \nonumber 
C_2=\left(
\begin{array}{cccccc}
 -1 & c_{12} & c_{13} & c_{12} c_{15}+c_{13} c_{16} &
   c_{15} & c_{16} \\
 0 & 0 & 0 & c_{12} c_{25}+c_{13} c_{26} & c_{25} & c_{26}
   \\
 0 & 0 & 0 & c_{12} c_{35}+c_{13} c_{36} & c_{35} & c_{36}
   \\
 0 & 0 & 0 & -1 & 0 & 0 \\
 0 & \frac{c_{36}}{c_{25} c_{36}-c_{26} c_{35}} &
   \frac{1}{c_{26}-\frac{c_{25} c_{36}}{c_{35}}} & \gamma_{54}
    & \frac{c_{35}}{2} & \frac{c_{36}}{2} \\
 0 & \frac{c_{26}}{c_{26} c_{35}-c_{25} c_{36}} &
   \frac{c_{25}}{c_{25} c_{36}-c_{26} c_{35}} & \gamma_{64}&
   -\frac{c_{25}}{2} & -\frac{c_{26}}{2} \\
\end{array}
\right)
\end{equation}
where
$$ \gamma_{54}=\frac{-2 c_{16} c_{35}+c_{12} \left(c_{25}
   c_{36}-c_{26} c_{35}\right) c_{35}+c_{36} \left(2
   c_{15}+c_{13} \left(c_{25} c_{36}-c_{26}
   c_{35}\right)\right)}{2 c_{25} c_{36}-2 c_{26} c_{35}},
  $$
$$ \gamma_{64}=\frac{-2
   c_{16} c_{25}+2 c_{15} c_{26}+\left(c_{12}
   c_{25}+c_{13} c_{26}\right) \left(c_{25} c_{36}-c_{26}
   c_{35}\right)}{2 c_{26} c_{35}-2 c_{25} c_{36}} .
  $$

For $c_{12} =c_{13}=c_{15} =c_{16} =c_{23} =c_{32}=0$ and $c_{22} =c_{33}=1$ the matrix $C_1$ is equal to $\beta$-shift
\begin{equation} \nonumber 
C_{10} = \left(
\begin{array}{cccccc}
 1 & 0 & 0 & 0 & 0 & 0 \\
 0 & 1 & 0 & 0 & 0 & 0 \\
 0 & 0 & 1 & 0 & 0 & 0 \\
 0 & 0 & 0 & 1 & 0 & 0 \\
 0 & 0 & \frac{1}{2} & 0 & 1 & 0 \\
 0 & -\frac{1}{2} & 0 & 0 & 0 & 1 \\
\end{array}
\right).
\end{equation}
Background
\begin{equation}\label{mtz52beta}
  \wh\cf(t,\hat x_1) =  \left(
\begin{array}{cccc}
 -1 & 0 & 0 & 0 \\
 0 & t^2 & 0 & 0 \\
 0 & 0 & \frac{4 e^{2 \hat x_1} t^2}{e^{4 \hat x_1} t^4+4} & \frac{2 e^{4 \hat x_1} t^4}{e^{4 \hat x_1} t^4+4} \\
 0 & 0 & -\frac{2 e^{4 \hat x_1} t^4}{e^{4 \hat x_1} t^4+4} & \frac{4 e^{2 \hat x_1} t^2}{e^{4 \hat x_1} t^4+4} \\
\end{array}
\right)
\end{equation}
obtained by this $\beta$-shift can be brought to the Brinkmann form of a plane parallel wave
\begin{equation}\label{ppw22}
ds^2=\frac{2 u^2 \left(u^4-5\right) \left(z_3^2+z_4^2\right)}{\left(u^4+1\right)^2}du^2+2du\,dv+d z_3^2+d z_4^2
\end{equation}
accompanied by a non-trivial torsion
\begin{equation}\label{ppw22tor}
\wh H = \frac{4u}{1+u^4}du\wedge dz_3 \wedge dz_4.
\end{equation}
This \bkg\ was found in \cite{hpp} as non-Abelian T-dual of flat metric. Dilaton that together with \eqref{mtz52beta}
satisfies \vbe {} is
$$
\hat \Phi(t,\hat x_1)= -\frac{1}{2} \ln \left( \frac{t^4 e^{4 \hat x_1}}{4}+1\right).
$$

Dual of \eqref{mtz52beta} calculated using $D_0 \cdot C_{10}$ reads
\begin{align}\label{mtz52fd}
& \wb\cf(t,\bar x_2,\bar x_3)= \\ & \nonumber \left(
\begin{array}{cccc}
 -1 & 0 & 0 & 0 \\
 0 & \frac{t^2}{t^4+\bar x_2^2+\bar x_3^2} & \frac{\bar x_3 t^2+2 \bar x_2}{2 \left(t^4+\bar x_2^2+\bar x_3^2\right)} & \frac{\bar x_2 t^2+2\bar  x_3}{2 \left(t^4+\bar x_2^2+\bar x_3^2\right)} \\
 0 & \frac{t^2 \bar x_3-2 \bar x_2}{2 \left(t^4+\bar x_2^2+\bar x_3^2\right)} & \frac{\left(\bar x_3^2+4\right) t^4+4 \bar x_3^2}{4 t^2
   \left(t^4+\bar x_2^2+\bar x_3^2\right)} & \frac{-2 t^6+\bar x_2 \bar x_3 t^4-4 \bar x_2^2 t^2-4 \bar x_2 \bar x_3}{4 t^2 \left(t^4+\bar x_2^2+\bar x_3^2\right)} \\
 0 & \frac{t^2 \bar x_2-2 \bar x_3}{2 \left(t^4+\bar x_2^2+\bar x_3^2\right)} & \frac{2 t^6+\bar x_2 \bar x_3 t^4+4 \bar x_2^2 t^2-4 \bar x_2 \bar x_3}{4 t^2
   \left(t^4+\bar x_2^2+\bar x_3^2\right)} & \frac{\left(\bar x_2^2+4\right) t^4+4 \bar x_2^2}{4 t^2 \left(t^4+\bar x_2^2+\bar x_3^2\right)} \\
\end{array}
\right),
\end{align}
and with dilaton
$$
\wb\Phi(t,\bar x_2,\bar x_3)=-\frac{1}{2} \ln \left(t^2 \left(t^4+\bar x_2^2+\bar x_3^2\right)\right)
$$
it satisfies Generalized Supergravity Equations with $\wb{\mathcal J}=(0,2,0,0)$. However, $Div\, \Theta= \left(0,\frac{8}{4+t^4},0,0\right)$.

For $c_{12} =c_{13}=c_{15} =c_{16} =c_{26}=c_{35}=0$ and $c_{25} =c_{36}=1$ the matrix $C_2$ is equal to $\beta$-shift followed by factorized duality
\begin{equation}\label{FD}
C_{20}=\left(
\begin{array}{cccccc}
 -1 & 0 & 0 & 0 & 0 & 0 \\
 0 & 0 & 0 & 0 & 1 & 0 \\
 0 & 0 & 0 & 0 & 0 & 1 \\
 0 & 0 & 0 & -1 & 0 & 0 \\
 0 & 1 & 0 & 0 & 0 & \frac{1}{2} \\
 0 & 0 & 1 & 0 & -\frac{1}{2} & 0 \\
\end{array}
\right).
\end{equation}

Background  obtained by transformation \eqref{FD} is given by tensor
\begin{equation}\label{mtz52fd3}
  \wh\cf(t,\hat x_1) =  \left(
\begin{array}{cccc}
 -1 & 0 & 0 & 0 \\
 0 & t^2 & 0 & 0 \\
 0 & 0 & \frac{4 e^{2 \hat x_1} t^2}{4 t^4+e^{4 \hat x_1}} & \frac{2
   e^{4 \hat x_1}}{4 t^4+e^{4 \hat x_1}} \\
 0 & 0 & -\frac{2 e^{4 \hat x_1}}{4 t^4+e^{4 \hat x_1}} & \frac{4
   e^{2 \hat x_1} t^2}{4 t^4+e^{4 \hat x_1}} \\
\end{array}
\right)
\end{equation}
that  again gives plane parallel wave  \eqref{ppw22} with torsion \eqref{ppw22tor}.
Dilaton that together with \eqref{mtz52fd3} satisfy \vbe {} is
$$
\wh\Phi(t,\hat x_1)=-\frac{1}{2} \ln \left(t^4 e^{-4\hat x_1}+\frac{1}{4}\right).
$$

Dual of \eqref{mtz52fd3} found from $D_0 \cdot C_{20}$ is
\begin{align}\label{mtz52fd2}
& \wb\cf(t,\bar x_2,\bar x_3)= \\ & \nonumber \left(
\begin{array}{cccc}
 -1 & 0 & 0 & 0 \\
 0 & \frac{1}{t^2 \left(\bar x_2^2+\bar x_3^2+1\right)} & \frac{2 \bar x_2 t^2+\bar x_3}{2 t^2 \left(\bar x_2^2+\bar x_3^2+1\right)} & \frac{2 \bar x_3 t^2+\bar x_2}{2 t^2
   \left(\bar x_2^2+\bar x_3^2+1\right)} \\
 0 & \frac{\bar x_3-2 t^2 \bar x_2}{2 t^2 \left(\bar x_2^2+\bar x_3^2+1\right)} & \frac{4 \left(\bar x_3^2+1\right) t^4+\bar x_3^2}{4 t^2
   \left(\bar x_2^2+\bar x_3^2+1\right)} & \frac{-4 \bar x_2 x_3 t^4-2 \left(2 \bar x_2^2+1\right) t^2+\bar x_2 \bar x_3}{4 t^2 \left(\bar x_2^2+\bar x_3^2+1\right)} \\
 0 & \frac{\bar x_2-2 t^2 \bar x_3}{2 t^2 \left(\bar x_2^2+\bar x_3^2+1\right)} & \frac{-4 \bar x_2 \bar x_3 t^4+\left(4 \bar x_2^2+2\right) t^2+\bar x_2 \bar x_3}{4 t^2
   \left(\bar x_2^2+\bar x_3^2+1\right)} & \frac{4 \left(\bar x_2^2+1\right) t^4+\bar x_2^2}{4 t^2 \left(\bar x_2^2+\bar x_3^2+1\right)} \\
\end{array}
\right).
\end{align}
It is torsionless and with dilaton
$$
\wb\Phi(t,\bar x_2,\bar x_3)=-\frac{1}{2} \ln \left(t^2\left(\bar x_2^2+\bar x_3^2+1\right)\right)
$$
it satisfies \vbe, i.e.  Generalized Supergravity Equations with $\wb{\mathcal J}=(0,0,0,0)$. However, $Div\, \Theta= \left(0,\frac{8t^4}{1+4t^4},0,0\right)$.

\subsubsection{Transformation of  $\mathfrak b_{V}\bowtie \mathfrak a$ to $\mathfrak b_{VI_{-1}}\bowtie \mathfrak b_{Vii}$ and to its dual}

Lie algebra $\mathfrak{d}= \mathfrak b_{VI_{-1}}\bowtie \mathfrak b_{Vii}$ is
spanned by basis $(\wh T_1,\wh T_2,\wh T_3,\wb T^1,\wb T^2,\wb T^3)$ with
algebraic relations
\begin{align}
\nonumber [ \wh T_1, \wh T_2] & =-\wh T_2, & [ \wh T_1, \wh T_3] & = \wh T_3,\\
[ \wh T_1, \wb T^1] & = -\wh T_2, & [ \wh T_1, \wb T^2] & = \wh T_1 +\wb T^2, & [ \wh T_1, \wb T^3] & = -\wb T^3,\\
\nonumber [ \wh T_2, \wb T^2] & = -\wb T^1, & [ \wh T_3, \wb T^2] & = \wh T_3, & [ \wh T_3, \wb T^3] & = -\wh T_2 +\wb T^1,\\
\nonumber [ \wb T^1, \wb T^2] & = -\wb T^1, & [ \wb T^2, \wb T^3] & = \wb T^3.
\end{align}
Left-invariant vector fields of $\wh\G$ and $\bar\G$ are
\begin{align*}
\wh V_1 & =\partial_{\hat x_1}+\hat x_2\,\partial_{\hat x_2}-\hat x_3\,\partial_{\hat x_3}, & \wh V_2 & =\partial_{\hat x_2}, & \wh V_3 & =\partial_{\hat x_3},\\
\bar V_1 & =\e^{\bar x_2}\partial_{\bar x_1}, & \bar V_2 & =\partial_{\bar x_2}-\,\bar x_3\partial_{\bar x_3}, & \bar V_3 & =\partial_{\bar x_3}.
\end{align*}

Examples of mappings $C$ that transform the algebraic structure of Manin triple
$\mathfrak
b_{VI_{-1}}\bowtie \mathfrak a$ to $\mathfrak
b_{VI_{-1}}\bowtie \mathfrak b_{Vii}$ are given by matrices
\begin{equation} \nonumber 
C_1=\left(
\begin{array}{cccccc}
 -1 & 0 & 0 & 0 & 0 & 0 \\
 0 & 1 & 0 & 0 & 0 & 0 \\
 0 & 0 & 0 & 0 & 0 & 1 \\
 0 & 1 & 0 & -1 & 0 & 0 \\
 1 & 0 & 0 & 0 & 1 & 0 \\
 0 & 0 & 1 & 0 & 0 & 0 \\
\end{array}
\right),
\end{equation}
\begin{equation} \nonumber 
C_2=\left(
\begin{array}{cccccc}
 1 & 0 & 0 & 0 & 0 & 0 \\
 0 & 0 & 0 & 0 & 1 & 0 \\
 0 & 0 & 1 & 0 & 0 & 0 \\
 0 & 0 & 0 & 1 & 1 & 0 \\
 -1 & 1 & 0 & 0 & 0 & 0 \\
 0 & 0 & 0 & 0 & 0 & 1 \\
\end{array}
\right).
\end{equation}

Background obtained by $C_1$ is given by tensor
\begin{equation}\label{mtz51615ii1}
  \wh\cf(t,\hat x_1,\hat x_3) = \left(
\begin{array}{cccc}
 -1 & 0 & 0 & 0 \\
 0 & \frac{e^{2 \hat x_1} \left(\hat x_3^2+1\right) t^2}{t^4+e^{2 \hat x_1} \left(\hat x_3^2+1\right)} & \frac{t^4}{t^4+e^{2 \hat x_1} \left(\hat x_3^2+1\right)} & \frac{e^{2 \hat x_1} \hat x_3 t^2}{t^4+e^{2 \hat x_1} \left(\hat x_3^2+1\right)} \\
 0 & -\frac{t^4}{t^4+e^{2 \hat x_1} \left(\hat x_3^2+1\right)} & \frac{t^2}{t^4+e^{2 \hat x_1} \left(\hat x_3^2+1\right)} & \frac{e^{2 \hat x_1} \hat x_3}{t^4+e^{2 \hat x_1} \left(\hat x_3^2+1\right)} \\
 0 & \frac{e^{2 \hat x_1} \hat x_3 t^2}{t^4+e^{2 \hat x_1} \left(\hat x_3^2+1\right)} & -\frac{e^{2 \hat x_1} \hat x_3}{t^4+e^{2 \hat x_1} \left(\hat x_3^2+1\right)} & \frac{e^{2 \hat x_1} \left(t^4+e^{2 \hat x_1}\right)}{t^2 \left(t^4+e^{2 \hat x_1} \left(\hat x_3^2+1\right)\right)} \\
\end{array}
\right).
\end{equation}
Together with the dilaton
$$
\wh\Phi(t,\hat x_1,\hat x_3)=-\frac{1}{2} \ln \left(t^6 e^{-4\hat x_1}+t^2 e^{-2 {\hat x_1}}\left(\hat x_3^2+1\right)\right)
$$
they satisfy \sugra s with $\hJ =(0,0,2,0)$, but $Div\,\Theta=(0,0,1,0)$.

Dual of \eqref{mtz51615ii1}
\begin{align}
\nonumber & \wb\cf(t,\bar x_1,\bar x_2,\bar x_3)=\\
&\left(
\begin{array}{cccc}
 -1 & 0 & 0 & 0 \\
 0 & \frac{t^2}{\Delta} & -\frac{\bar x_1 t^2+e^{\bar x_2} \left(t^4+1\right)-1}{\Delta} & \frac{e^{3 \bar x_2} \bar x_3 t^4}{\Delta} \\
 0 & \frac{-\bar x_1 t^2+e^{\bar x_2} \left(t^4+1\right)-1}{\Delta} & \frac{t^2 \left(\bar x_1^2+e^{4 \bar x_2} \bar x_3^2 \left(t^4+1\right)+1\right)}{\Delta} & \frac{e^{3 \bar x_2} \bar x_3 t^2 \left(-\bar x_1 t^2+e^{\bar x_2} \left(t^4+1\right)-1\right)}{\Delta} \\
 0 & -\frac{e^{3 \bar x_2} \bar x_3 t^4}{\Delta} & \frac{e^{3 \bar x_2} \bar x_3 t^2 \left(\bar x_1 t^2+e^{\bar x_2} \left(t^4+1\right)-1\right)}{\Delta} & \frac{e^{2 \bar x_2} t^2 \left(e^{2 \bar x_2} \left(t^4+1\right)-2 e^{\bar x_2}+1\right)}{\Delta} \\
\end{array}
\right)
\end{align}
where
$$
\Delta = e^{4 \bar x_2} \bar x_3^2 t^4-2 e^{\bar x_2}+e^{2 \bar x_2} \left(t^4+1\right)+1
$$
and dilaton
\begin{equation}
\wb\Phi(t,\bar x_2,\bar x_3)=-\frac{1}{2} \ln \Delta
\end{equation}
calculated by the formula \eqref{dualdil} satisfy Generalized Supergravity Equations with vector $\bJ = (0,-\e^{\bar x_2},0,0)$  but $\ Div\,\Theta=\left(0,\frac{1}{t^4+1}-e^{\bar x_2},0,0\right)$.

Background obtained by $C_2$ is torsionless and has the form
\begin{equation}\label{mtz51615ii2}
  \wh\cf(t,\hat x_1,\hat x_3) =\left(
\begin{array}{cccc}
 -1 & 0 & 0 & 0 \\
 0 & \frac{e^{2 \hat x_1} \left(\hat x_3^2+1\right) t^2}{e^{2 \hat x_1} \left(\hat x_3^2+1\right)+1} & \frac{1}{e^{2 \hat x_1} \left(\hat x_3^2+1\right)+1} & \frac{e^{2 \hat x_1} \hat x_3 t^2}{e^{2 \hat x_1} \left(\hat x_3^2+1\right)+1} \\
 0 & -\frac{1}{e^{2 \hat x_1} \left(\hat x_3^2+1\right)+1} & \frac{1}{\left(e^{2 \hat x_1} \left(\hat x_3^2+1\right)+1\right) t^2} & \frac{e^{2 \hat x_1} \hat x_3}{e^{2 \hat x_1} \left(\hat x_3^2+1\right)+1} \\
 0 & \frac{e^{2 \hat x_1} \hat x_3 t^2}{e^{2 \hat x_1} \left(\hat x_3^2+1\right)+1} & -\frac{e^{2 \hat x_1} \hat x_3}{e^{2 \hat x_1} \left(\hat x_3^2+1\right)+1} & \frac{e^{2 \hat x_1} \left(1+e^{2 \hat x_1}\right) t^2}{e^{2 \hat x_1} \left(\hat x_3^2+1\right)+1} \\
\end{array}
\right).
\end{equation}
Together with dilaton
$$
\wh\Phi(t,\hat x_1,\hat x_3)=-\frac{1}{2} \ln \left(t^2 \left(1+e^{2 {\hat x_1}} \left(\hat x_3^2+1\right)\right)\right)
$$
it satisfies Generalized Supergravity Equations with vector $\hJ =(0,0,0,0)$, but $Div\,\Theta=(0,0,1,0)$.

Dual of \eqref{mtz51615ii2} given by
\begin{align}
&\nonumber \wb\cf(t,\bar x_1,\bar x_2,\bar x_3)=\\
\nonumber 
 &\left(
\begin{array}{cccc}
 -1 & 0 & 0 & 0 \\
 0 & \frac{t^2}{\Delta} & \frac{\left(1-2 e^{\bar x_2}\right) t^4-t^2
   {\bar x_1}}{\Delta} & \frac{e^{3 \bar x_2} \bar x_3}{\Delta} \\
 0 & \frac{\left(-1+2 e^{\bar x_2}\right) t^4-t^2 {\bar x_1}}{\Delta} & \frac{t^2
   \left(t^4+{\bar x_1}^2+2 e^{4 \bar x_2} \bar x_3^2\right)}{\Delta} & \frac{e^{3 \bar x_2}
   \left(\left(-1+2 e^{\bar x_2}\right) t^2-{\bar x_1}\right) \bar x_3}{\Delta} \\
 0 & -\frac{e^{3 \bar x_2} \bar x_3}{\Delta} & \frac{e^{3 \bar x_2} \left(\left(-1+2
   e^{\bar x_2}\right) t^2+{\bar x_1}\right) \bar x_3}{\Delta} & \frac{e^{2 \bar x_2} \left(1-2
   e^{\bar x_2}+2 e^{2 \bar x_2}\right) t^2}{\Delta} \\
\end{array}
\right)
\end{align}
where
$$
\Delta = \left(1-2 e^{\bar x_2}+2 e^{2 \bar x_2}\right) t^4+e^{4 \bar x_2} \bar x_3^2
$$
and dilaton
\begin{equation} \nonumber
\wb\Phi(t,\bar x_2,\bar x_3)=-\frac{1}{2} \ln \left(t^4\left(e^{-4 {\bar x_2}}-2 e^{-3{\bar x_2}}+2 e^{-2 \bar x_2}\right)+\bar x_3^2\right)
\end{equation}
calculated by formula \eqref{dualdil} satisfy Generalized Supergravity Equations with vector $\bJ = (0,e^{\bar x_2},0,0)$. On the other hand, $Div\,\Theta=\left(0,\frac{1}{2}-e^{\bar x_2},0,0\right)$.

\section{\pltp y of curved cosmologies}

\subsection{Bianchi $VI_{-1}$ cosmology}\label{B_VI-1}

Next we will transform curved Bianchi $VI_{-1}$ cosmology given by metric
\begin{equation} \nonumber 
\cf(t,y_1)=\left(
\begin{array}{cccc}
 -e^{-4 \Phi(t)}a_1(t)^2{a_2}(t)^4 & 0 & 0 & 0 \\
 0 &a_1(t)^2 & 0 & 0 \\
 0 & 0 & e^{-2 y_1}{a_2}(t)^2 & 0 \\
 0 & 0 & 0 & e^{2 y_1}{a_2}(t)^2 \\
\end{array}
\right)
\end{equation}
where
\begin{equation} \nonumber
a_1(t) = \sqrt{p_1} \exp \left(\frac{1}{2} e ^{2 p_2 t} + \frac{p_1 t}{2} +\Phi(t)\right), \qquad a_2(t) = \sqrt{p_2} e^{\frac{p_2 t}{2}+ \Phi(t)}
\end{equation}
and dilaton $\Phi = \beta t$.
Its scalar curvature is
$$
R=\frac{\left(-12 \beta ^2+2 p_1 p_2+p_2^2\right) e^{-t (p_1+2 (\beta +p_2))-e^{2 p_2 t}}}{2 p_1 p_2^2}
$$
and \vbe\ reduce to condition
$$
\beta^2=\frac{1}{4}(2 p_1 p_2 + p_2^2)
$$
for constants $p_1, p_2$ and $\beta$.

The metric is invariant \wrt\ the action of Bianchi $VI_{_{-1}}$ group and can be
constructed by virtue of Manin triple $\mathfrak{d}=\mathfrak b_{VI_{-1}}\bowtie \mathfrak a$ spanned by basis $(T_1,T_2,T_3,\wwt T^1,\wwt T^2,\wwt T^3)$ with algebraic relations
\begin{align}
[T_1, T_2] & = - T_2, & [  T_1,  T_3] & =  T_3, & [  T_1, \wwt T^2] & = \wwt T^2,\\ \nonumber
[  T_1, \wwt T^3] & = -\wwt T^3, & [ T_2, \wwt T^2] & = -\wwt T^1, &[  T_3, \wwt T^3] & = \wwt T^1.
\end{align}
Structure constants of $\mathfrak b_{VI_{-1}}$ are traceless.

{\dd\ is the same as in Section \ref{secB5}, where we have seen that beside $\mathfrak{d}=\mathfrak b_{VI_{-1}}\bowtie \mathfrak a$ the algebra $\cd$ can be decomposed into Manin triples $\mathfrak b_{V}\bowtie \mathfrak a$, $\mathfrak b_{V}\bowtie \mathfrak b_{II}$, $\mathfrak b_{VI_{-1}}\bowtie \mathfrak b_{V.ii}$ and their duals. In this Section, however, the \bkg s are different as we use different matrix $E(s)$.}

Formula \eqref{dualdil} for new dilatons
\begin{equation} \nonumber
\wh\Phi(t,\hat x)= \beta\,t-\half\ln \Big| \det \left[\left(N + \wh\Pi(\hat x) M\right)\wh a(\hat x)\right]\Big|
\end{equation}
does not depend on \coor s $y$ and is, therefore, applicable for any Manin triple of this \dd.

\subsubsection{Transformation of  $\mathfrak b_{VI_{-1}}\bowtie \mathfrak a$ to $\mathfrak b_{V}\bowtie \mathfrak a$ and to its dual}

Lie algebra $\mathfrak{d}=\mathfrak b_{V}\bowtie \mathfrak a$ is
spanned by $(\wh T_1,\wh T_2,\wh T_3,\wb T^1,\wb T^2,\wb T^3)$ with
the algebraic relations
\begin{align}
[\wh T_1, \wh T_2] & = \wh T_2, & [ \wh T_1, \wh T_3] & = \wh T_3, & [ \wh T_1, \wb T^2] & = -\wb T^2,\\ \nonumber
[ \wh T_1, \wb T^3] & = -\wb T^3, & [\wh T_2, \wb T^2] & = \wb T^1, &[ \wh T_3, \wb T^3] & = \wb T^1.
\end{align}
Left-invariant vector fields of $\wh\G$ and $\bar\G$ are
\begin{align*}
\wh V_1 & =\partial_{\hat x_1}-\hat x_2\,\partial_{\hat x_2}-\hat x_3\,\partial_{\hat x_3}, & \wh V_2 & =\partial_{\hat x_2}, & \wh V_3 & =\partial_{\hat x_3},\\
\bar V_1 & =\partial_{\bar x_1}, & \bar V_2 & =\partial_{\bar x_2}, & \bar V_3 & =\partial_{\bar x_3}.
\end{align*}

Mappings $C$ that transform the algebraic structure of Manin triple $\mathfrak b_{VI_{-1}}\bowtie \mathfrak a$ to $\mathfrak b_{V}\bowtie \mathfrak a$ are given by matrices inverse to \eqref{C6151} and
\eqref{C6151b}. After a suitable change of coordinates we find that backgrounds obtained from these general solutions differ from \bkg\
\begin{equation}\label{mtz51fd1}
  \wh\cf(t,\hat x_1) =  \left(
\begin{array}{cccc}
 -e^{-4 t \beta } a_1(t)^2 a_2(t)^4 & 0 & 0 & 0 \\
 0 & a_1(t)^2 & 0 & 0 \\
 0 & 0 & e^{2 \hat x_1} a_2(t)^2 & 0 \\
 0 & 0 & 0 & \frac{e^{2 \hat x_1}}{a_2(t)^2} \\
\end{array}
\right)
\end{equation}
obtained using factorized duality \eqref{FD4} only by a torsionless $\wh{\mathcal{B}}$ field. Note that $\wh{\cf}$ is again invariant \wrt\ group $\mathscr B_{V}$. This metric is not flat and dilaton that together with \eqref{mtz51fd1} satisfy \vbe {} is
$$
\wh\Phi(t,\hat x_1)=\beta  t+\hat x_1 - \ln a_2(t).
$$

Dual of \eqref{mtz51fd1} is given by tensor
\begin{align} \label{mtz15fd1}
&\wb\cf(t,\bar x_2,\bar x_3)=\\ \nonumber &\left(
\begin{array}{cccc}
 -e^{-4 \beta t} a_1(t)^2 a_2(t)^4 & 0 & 0 & 0 \\
 0 & \frac{a_2(t)^2}{\Delta} & \frac{\bar x_2}{\Delta} & \frac{a_2(t)^4 \bar x_3}{\Delta} \\
 0 & -\frac{\bar x_2}{\Delta} & \frac{a_1(t)^2+a_2(t)^2 \bar x_3^2}{\Delta} & -\frac{a_2(t)^2 \bar x_2 \bar x_3}{\Delta} \\
 0 & -\frac{a_2(t)^4 \bar x_3}{\Delta} & -\frac{a_2(t)^2 \bar x_2 \bar x_3}{\Delta} & \frac{a_2(t)^2 \left(a_1(t)^2 a_2(t)^2+\bar x_2^2\right)}{\Delta} \\
\end{array}
\right)
\end{align}
where
$$
\Delta=\bar x_3^2 a_2(t)^4+a_1(t)^2 a_2(t)^2+\bar x_2^2.
$$
Together with dilaton
$$
\wb\Phi(t,\bar x_2,\bar x_3)=\beta t - \frac{1}{2} \ln \Delta
$$
\bkg\ \eqref{mtz15fd1} satisfies the Generalized Supergravity Equations \eqref{betaG}-\eqref{betaPhi} for $\wb{\mathcal J} =(0,1,0,0)$. On the other hand $Div\,\Theta=(0,2,0,0)$.

\subsection{Bianchi $VI_\kappa$ cosmology}\label{secB6k}

Now we are ready to study \pltp ies of the most complicated curved cosmology invariant \wrt {} Bianchi $VI_\kappa$. Its Lie algebra is contained in semi-Abelian six-dimensional Manin triple\footnote{Linear \tfn\ between basis of $\mathfrak b_{VI_\kappa}\bowtie \mathfrak a$ used in this paper and vectors $X_i, \wwt X_j$ that span Lie algebra of \dd\ $(6_a|1)$ in \cite {snohla:ddoubles} is
\begin{equation}\label{6ato6kappa}
T_1=-\frac{1}{1+a}X_1, \quad T_2=X_2- X_3,\quad T_3= X_2+X_3,\quad \kappa=\frac{a-1}{a+1},\ a>0.
\end{equation}}
$\mathfrak{d}=\mathfrak b_{VI_\kappa}\bowtie \mathfrak a$ spanned by
basis $(T_1,T_2,T_3,\wwt T^1,\wwt T^2,\wwt T^3)$ with non-trivial algebraic
relations
\begin{align}
\label{comB6}
[T_1, T_2]&=\kappa\, T_2, & [ T_1, T_3]& = T_3, & & \kappa\neq -1, \\
\label{MT6kappa1} [ T_1, \wwt T^2] & =-\kappa\,\wwt T^2, & [ T_1, \wwt T^3]& =-\wwt T^3, & [ T_2, \wwt T^2] & = \kappa\,\wwt T^1, & [ T_3, \wwt T^3]= \wwt T^1.
\end{align}
Note that for $\kappa = 0$, or $\kappa = 1$,
these are \comrel s of $\mathfrak b_{III}$, or $\mathfrak b_{V}$
respectively. The case $\kappa = -1$ was treated separately in
section \ref{B_VI-1}. The group ${\B}_{VI_\kappa}$ is not semisimple and trace of its
structure constants does not vanish. Lie algebra of the \dd {} $({\B}_{VI_\kappa}|\A)$
admits several other Manin triples \cite{snohla:ddoubles}.

Metric of Bianchi ${VI_\kappa}$ cosmology reads \cite{hokico}
\begin{equation} \label{mtz6kappa1}
\cf(t,y_1)=\left(
\begin{array}{cccc}
 -e^{-4 \Phi(t)}a_1(t)^2{a_2}(t)^2{a_3}(t)^2 & 0 & 0 & 0 \\
 0 &a_1(t)^2 & 0 & 0 \\
 0 & 0 & e^{2\kappa y_1}{a_2}(t)^2 & 0 \\
 0 & 0 & 0 & e^{2 y_1}{a_3}(t)^2 \\
\end{array}
\right)
\end{equation}
where the functions $a_i(t)$ have the form
\begin{align}\label{B6k_ai}
{a_1}(t)&=e^{\Phi(t)}\left(\frac{{p_1}}{\kappa +1}\right)^{\frac{\kappa^2+1}{(\kappa +1)^2}}e^{\frac{(\kappa -1) {p_2} t}{2 (\kappa +1)}}\sinh ^{-\frac{\kappa ^2+1}{(\kappa+1)^2}}({p_1} t),\nonumber\\
{a_2}(t)&= e^{\Phi(t)}\left(\frac{{p_1}}{\kappa +1}\right)^{\frac{\kappa}{\kappa +1}} e^{\frac{{p_2} t}{2}} \sinh ^{-\frac{\kappa}{\kappa +1}}({p_1} t),\\
{a_3}(t)&= e^{\Phi(t)}\left(\frac{{p_1}}{\kappa
+1}\right)^{\frac{1}{\kappa +1}}e^{-\frac{{p_2} t}{2}} \sinh
^{-\frac{\kappa}{\kappa +1}}({p_1} t),\nonumber
\end{align}
and dilaton is $\Phi(t) = \beta t$.
The background \eqref{mtz6kappa1} is invariant with respect to symmetry generated by
left-invariant vector fields
\begin{equation} \nonumber 
V_1=\partial_{y_1}-\kappa\,y_2\,\partial_{y_2}-y_3\partial_{y_3},
\qquad V_2=\partial_{y_2}, \qquad V_3=\partial_{y_3}
\end{equation}
satisfying \eqref{comB6}. The \vbe\ reduce to condition
\begin{equation}\label{betapcond}
\beta^2=\frac{ \left(\kappa^2+\kappa +1\right) {p_1}^2}{(\kappa
+1)^2}-\frac{{p_2}^2}{4}.
\end{equation}
The background is torsionless and for $\beta=0$ also Ricci flat.

Formula \eqref{dualdil} for new dilatons then reads
\begin{equation} \nonumber
\wh\Phi(\hat x, \bar x)=\beta\,t-\half(1+\kappa)y_1(\hat x, \bar x)-\half\ln \Big| \det \left[\left(N + \wh\Pi(\hat x) M\right)\wh a(\hat x)\right]\Big|
\end{equation}
so that, it is again necessary to solve $y_1$
from \eqref{lgh} and find how it depends on $\hat x$ and $\bar x$.

\subsubsection{Transformation of $\mathfrak b_{VI_\kappa}\bowtie \mathfrak a$ to $\mathfrak b_{VI_\kappa}\bowtie \mathfrak b_{II}$ and to its dual}

Manin triple $\mathfrak{d}=\mathfrak b_{VI_\kappa}\bowtie \mathfrak b_{II}$, which belongs to decompositions of the Lie algebra of $(\mathscr B_{VI_\kappa}|\mathscr A)$, is spanned by $(\wh T_1,\wh T_2,\wh T_3,\wb T^1,\wb T^2,\wb T^3)$ with algebraic relations
\begin{align}
\nonumber [\wh T_1, \wh T_2] & =\kappa \wh T_2, & [ \wh T_1, \wh T_3] & = \wh T_3, & &\\
\label{MT6kappa2} [ \wh T_1, \wb T^2] & =\wh T_3-\kappa\,\wb T^2, & [ \wh T_1, \wb T^3] & =-\wh T_2-\wb T^3,\\
\nonumber [ \wh T_2, \wb T^2] & = \kappa\,\wb T^1, & [ \wh T_3, \wb T^3] & = \wb T^1, & [\wb T^2,  \wb T^3]& = \wb T^1.
\end{align}
Left-invariant fields of $\wh\G$ and $\bar\G$ are
\begin{align*}
\wh V_1 & =\partial_{\hat x_1}-\kappa\, \hat x_2\,\partial_{\hat x_2}- \hat x_3\,\partial_{\hat x_3}, & \wh V_2 & =\partial_{\hat x_2}, & \wh V_3 & =\partial_{\hat x_3}, \\
\bar V_1 & =\partial_{\bar x_1}, &  \bar V_2 & =-\bar x_3\,\partial_{\bar x_1}+\,\partial_{\bar x_2}, & \bar V_3 & =\partial_{\bar x_3}.
\end{align*}

There are two different linear mappings \eqref{C_mat} between  $\mathfrak
b_{VI_\kappa}\bowtie \mathfrak a$ and $\mathfrak
b_{VI_\kappa}\bowtie \mathfrak b_{II}$.  One of them is given by the
matrix
\begin{equation} \nonumber 
C_1=\left(
\begin{array}{cccccc}
 -1 & c_{12} & c_{13} & c_{12} c_{15}+c_{13} c_{16} & c_{15} & c_{16} \\
 0 & 0 & 0 & \frac{c_{12}}{c_{52}} & \frac{1}{c_{52}} & 0 \\
 0 & 0 & 0 & (\kappa +1) c_{13} c_{56} & 0 & (\kappa +1) c_{56} \\
 0 & 0 & 0 & -1 & 0 & 0 \\
 0 & c_{52} & 0 & c_{15} c_{52}+c_{13} c_{56} & 0 & c_{56} \\
 0 & 0 & \frac{1}{(\kappa + 1) c_{56}} & \frac{\frac{c_{16}}{c_{56}}-\frac{c_{12}}{c_{52}}}{\kappa +1} & -\frac{1}{(\kappa +1) c_{52}} & 0 \\
\end{array}
\right)
\end{equation}
and the second is
\begin{equation} \nonumber 
C_2=\left(
\begin{array}{cccccc}
 1 & c_{12} & c_{13} & -c_{12} c_{15}-c_{13} c_{16} & c_{15} & c_{16} \\
 0 & \frac{1}{c_{55}} & 0 & -\frac{c_{15}}{c_{55}} & 0 & 0 \\
 0 & 0 & (\kappa +1) c_{53} & -(\kappa +1) c_{16} c_{53} & 0 & 0 \\
 0 & 0 & 0 & 1 & 0 & 0 \\
 0 & 0 & c_{53} & -c_{16} c_{53}-c_{12} c_{55} & c_{55} & 0 \\
 0 & -\frac{1}{(\kappa +1) c_{55}} & 0 & \frac{\frac{c_{15}}{c_{55}}-\frac{c_{13}}{c_{53}}}{\kappa +1} & 0 & \frac{1}{(\kappa + 1)c_{53}} \\
\end{array}
\right).
\end{equation}
Tensors $\wh{\cal F}$ on $\wh\G=\G={\B}_{VI_\kappa}$ calculated from
these general forms are rather extensive. {Nevertheless, up to change of coordinates and gauge shift the \bkg s coming from $C_1$ are equivalent to
\begin{equation}
\label{hatFkappa21}
\wh \cf (t,\hat x_1) =
\left(
\begin{array}{cccc}
 -e^{-4 \beta t} a_1^2 a_2^2 a_3^2 & 0 & 0 & 0 \\
 0 & a_1^2 & 0 & 0 \\
 0 & 0 & \frac{e^{2 \kappa  \hat x_1} a_3^2}{a_2^2 a_3^2 c_{52}^2+e^{2 (\kappa +1) \hat x_1} c_{56}^2} & \frac{e^{2 (\kappa +1) \hat x_1} (\kappa +1) c_{56}^2}{a_2^2 a_3^2 c_{52}^2+e^{2 (\kappa +1) \hat x_1} c_{56}^2} \\
 0 & 0 & -\frac{e^{2 (\kappa +1) \hat x_1} (\kappa +1) c_{56}^2}{a_2^2 a_3^2 c_{52}^2+e^{2 (\kappa +1) \hat x_1} c_{56}^2} & \frac{e^{2 \hat x_1} (\kappa +1)^2 a_2^2 c_{52}^2 c_{56}^2}{a_2^2 a_3^2 c_{52}^2+e^{2 (\kappa +1) \hat x_1} c_{56}^2} \\
\end{array}
\right)
\end{equation}
that can be obtained from $C_1$ when all constants $c_{ij}$ except $c_{52}$ and $c_{56}$ are set to zero.}
Since $y_1=-\hat x_1$, we get
\begin{equation} \nonumber 
\wh\Phi(t,\hat x_1)=\beta t-\frac{1}{2} \ln \left(\frac{c_{52}^2 a_2(t)^2 a_3(t)^2 e^{-2 \hat x_1 (\kappa +1)}+c_{56}^2}{(\kappa +1) c_{52} c_{56}}\right)
\end{equation}
and $\mathcal{\wh J}= (0,0,0,0)=Div\,\Theta$. Dilaton together with \eqref{hatFkappa21} satisfies \vbe {}  if the condition \eqref{betapcond} holds.
Coordinate transformation brings background on $\bar\G={\B}_{II}$ calculated from $D_0\cdot C_1$ to the form
\begin{align}\nonumber 
& \wb \cf (t,\bar x_2,\bar x_3) = \\ \nonumber
& \left(
\begin{array}{cccc}
 -e^{-4 \beta t} a_1^2 a_2^2 a_3^2 & 0 & 0 & 0 \\
 0 & \frac{1}{\Delta} & \frac{(\kappa +1) a_2^2 \bar x_2}{\kappa  c_{56} \Delta} & \frac{a_3^2 \bar x_3}{(\kappa +1) c_{56} \Delta} \\
 0 & -\frac{(\kappa +1) a_2^2 \bar x_2}{\kappa  c_{56} \Delta} & \frac{a_2^2 \left((\kappa +1)^2 a_1^2 c_{56}^2+a_3^2 \bar x_3^2\right)}{\kappa ^2 c_{56}^2 \Delta} & \frac{-c_{56} \Delta - (\kappa +1) a_2^2 a_3^2 c_{52} \bar x_2 \bar x_3 }{\kappa  (\kappa +1) c_{52} c_{56}^2 \Delta} \\
 0 & -\frac{a_3^2 \bar x_3}{(\kappa +1) c_{56} \Delta} & \frac{c_{56}\Delta-(\kappa +1) a_2^2 a_3^2 c_{52} \bar x_2 \bar x_3}{\kappa  (\kappa +1) c_{52} c_{56}^2 \Delta} & \frac{a_3^2 \left(a_1^2 c_{56}^2+a_2^2 \bar x_2^2\right)}{c_{56}^2 \Delta} \\
\end{array}
\right)
\end{align}
where
$$
\Delta = (\kappa +1)^2 a_1^2 c_{56}^2+(\kappa +1)^2 a_2^2 \bar x_2^2+a_3^2 \bar x_3^2.
$$
With the corresponding dilaton
\begin{equation} \nonumber 
\wb\Phi(t,\bar x_2,\bar x_3)= \beta t-\frac{1}{2} \log \left(\frac{\Delta}{(\kappa +1) c_{52} c_{56}}\right)
\end{equation}
the \bkg\ $\wb \cf$ satisfies Generalized Supergravity
Equations with $\mathcal{\wb J}= (0,0,0,0)$ obtained from $y_1=-\hat x_1$
and \eqref{kilJbar}. However, $Div\, \Theta$ is nontrivial and depends on $a_2(t)a_3(t).$

Transformation of Manin triple given by $C_2$ gives similar results. All constants $c_{ij}$ except $c_{53}$ and $c_{55}$ can be set to zero, and the only relevant difference is that in this case
$y_1=\hat x_1$, so the Generalized Supergravity
Equations are satisfied with $\mathcal{\wh J}= (0,0,0,0)$ and
$\mathcal{\wb J}= (0,\kappa+1,0,0)$.

\subsubsection{Transformation of $\mathfrak b_{VI_\kappa}\bowtie \mathfrak a$ to $\mathfrak b_{VI_\kappa}\bowtie \mathfrak b_{VI_{-\kappa}.iii}$ and to its dual}

The Manin triple $\mathfrak b_{VI_\kappa}\bowtie \mathfrak b_{VI_{-\kappa}.iii}$ is another Manin triple of the \dd {}  $(\mathscr B_{VI_\kappa}|\mathscr A)$. It corresponds to $(6_a|6_{1/a}.iii)$ in \cite{snohla:ddoubles} by the \tfn {} \eqref{6ato6kappa}. It is spanned by $(\wh T_1,\wh T_2,\wh
T_3,\wb T^1,\wb T^2,\wb T^3)$ with algebraic relations
$$
[ \wh T_1, \wh T_2]=\kappa \wh T_2, \qquad [ \wh T_1, \wh T_3]= \wh T_3, \qquad [\wh T_1, \wb T^1]=\wh T_3, \qquad [ \wh T_1, \wb T^2]=-\kappa\,\wb T^2,
$$
\begin{equation}
\label{MT6kappaiii}[ \wh T_1, \wb T^3]=-\wh T_1-\wb T^3,\qquad [ \wh T_2, \wb T^2]=
\kappa(-\wh T_3+\wb T^1), \qquad [ \wh T_2, \wb T^3]= \kappa\,\wh T_2,
\end{equation}
$$
[ \wh T_3, \wb T^3]= \wb T^1,\qquad [\wb T^1, \wb T^3]= \wb T^1, \qquad [\wb T^2, \wb
T^3]=-\kappa\, \wb T^2.
$$
Left-invariant fields of $\wh\G$ and $\bar\G$ are
\begin{align*}
\wh V_1 & =\partial_{\hat x_1}-\kappa\, \hat x_2\,\partial_{\hat x_2}- \hat x_3\,\partial_{\hat x_3}, & \wh V_2 & =\partial_{\hat x_2}, & \wh V_3 & =\partial_{\hat x_3},\\
\bar V_1 & =\e^{-\bar x_3}\partial_{\bar x_1}, & \bar V_2 & =\e^{\kappa\,\bar x_3}\,\partial_{\bar x_2}, & \bar V_3 & =\partial_{\bar x_3}.
\end{align*}

There are two different \pltp ies between  $\mathfrak
b_{VI_\kappa}\bowtie \mathfrak a$ and $\mathfrak
b_{VI_\kappa}\bowtie \mathfrak b_{VI_{-\kappa}.iii}$.  One of them
is given by the matrix
\begin{equation} \nonumber 
C_1=\left(
\begin{array}{cccccc}
 -1 & c_{12} & c_{13} & c_{12} c_{15}+c_{13} c_{16} & c_{15} & c_{16} \\
 0 & 0 & 0 & \frac{c_{12}}{c_{52}} & \frac{1}{c_{52}} & 0 \\
 0 & 0 & 0 & c_{13} c_{36} & 0 & c_{36} \\
 0 & 0 & 0 & c_{13} c_{36}-1 & 0 & c_{36} \\
 0 & c_{52} & 0 & c_{15} c_{52} & 0 & 0 \\
 1 & -c_{12} & \frac{1}{c_{36}}-c_{13} & c_{16} \left(\frac{1}{c_{36}}-c_{13}\right)-c_{12} c_{15} & -c_{15} & -c_{16} \\
\end{array}
\right)
\end{equation} and the other one is
\begin{equation} \nonumber 
C_2=\left(
\begin{array}{cccccc}
 1 & c_{12} & c_{13} & -c_{12} c_{15}-c_{13} c_{16} & c_{15} &
   c_{16} \\
 0 & c_{22} & 0 & -c_{15} c_{22} & 0 & 0 \\
 0 & 0 & c_{33} & -c_{16} c_{33} & 0 & 0 \\
 0 & 0 & c_{33} & 1-c_{16} c_{33} & 0 & 0 \\
 0 & 0 & 0 & -\frac{c_{12}}{c_{22}} & \frac{1}{c_{22}} & 0 \\
 -1 & -c_{12} & -c_{13} & c_{12} c_{15}+c_{13}
   \left(c_{16}-\frac{1}{c_{33}}\right) & -c_{15} &
   \frac{1}{c_{33}}-c_{16} \\
\end{array}
\right).
\end{equation}

General forms of tensors and dilatons
$\wh{\cal F}, \wh\Phi$ on $\wh\G=\G={\B}_{VI_\kappa}$ and $\wb{\cal
F},\wb\Phi$ on $\wb\G={\B}_{VI_{-\kappa}.iii}$  are too extensive
to display. Setting $c_{12}=c_{13}=c_{15}=c_{16}=0$ and $c_{52}=c_{36}=1$ in $C_1$ we get
\begin{align} \nonumber
& \wh{\cf}(t,\hat x_1,\hat x_2)=\\ \nonumber & \left(
\begin{array}{cccc}
 -e^{-4 \beta t} a_1^2 a_2^2 a_3^2 & 0 & 0 & 0 \\
 0 & \frac{a_1^2 \left(a_2^2 a_3^2+e^{2 (\kappa +1) \hat x_1} \kappa ^2 \hat x_2^2\right)}{\Delta} & \frac{e^{2 (\kappa +1) \hat x_1} \kappa  a_1^2 \hat x_2}{\Delta} & \frac{e^{2 \hat x_1} a_1^2 a_2^2}{\Delta} \\
 0 & \frac{e^{2 (\kappa +1) \hat x_1} \kappa  a_1^2 \hat x_2}{\Delta} & \frac{e^{2 \kappa  \hat x_1} \left(e^{2 \hat x_1} a_1^2+a_3^2\right)}{\Delta} & -\frac{e^{2 (\kappa +1) \hat x_1} \kappa  \hat x_2}{\Delta} \\
 0 & -\frac{e^{2 \hat x_1} a_1^2 a_2^2}{\Delta} & \frac{e^{2 (\kappa +1) \hat x_1} \kappa  \hat x_2}{\Delta} & \frac{e^{2 \hat x_1} a_2^2}{\Delta} \\
\end{array}
\right)
\end{align}
where
$$\Delta = e^{2 \hat x_1} a_1^2 a_2^2+a_3^2 a_2^2+e^{2 (\kappa +1) \hat x_1} \kappa ^2 \hat x_2^2
$$
and
\begin{equation} \nonumber
\wh\Phi(t,\hat x_1,\hat x_2)=\beta  t-\frac{1}{2} \log \left(-e^{-2 (\kappa +1) {\hat x_1}} \Delta\right).
\end{equation}
Generalized Supergravity Equations with  $\wh{\mathcal
J}=(0,0,0,\kappa)$ obtained from
\eqref{kilJ} and $y_1=-\hat x_1+\bar x_3$ are satisfied under the condition
\eqref{betapcond}.

Dual \tfn\ gives
\begin{align} \nonumber
& \wb{\cf}(t,\bar x_1,\bar x_2,\bar x_3)=\\ \nonumber
& \left(
\begin{array}{cccc}
 \frac{-e^{4 t \beta }}{a_1^2 a_2^2 a_3^2} & 0 & 0 & 0 \\
 0 & \frac{a_1^2 a_3^2+\bar x_1^2}{a_1^2+a_3^2} & -\frac{\kappa  \left(a_3^2+\bar x_1\right)
   \bar x_2}{a_1^2+a_3^2} & \frac{e^{-\bar x_3} a_1^2+\left(e^{-\bar x_3}-1 \right) a_3^2-\bar x_1}{a_1^2+a_3^2} \\
 0 & \frac{\kappa  \left(a_3^2-\bar x_1\right) \bar x_2}{a_1^2+a_3^2} & \frac{a_1^2+a_3^2+\kappa
   ^2 a_2^2 \bar x_2^2}{a_2^2 \left(a_1^2+a_3^2\right)} & \frac{\kappa
   \bar x_2}{a_1^2+a_3^2} \\
 0 & -\frac{e^{-\bar x_3} a_1^2+\left(-1+e^{-\bar x_3}\right) a_3^2+\bar x_1}{a_1^2+a_3^2} & \frac{\kappa
   \bar x_2}{a_1^2+a_3^2} & \frac{1}{a_1^2+a_3^2} \\
\end{array}
\right)\-1
\end{align}
and
\begin{equation} \nonumber
\wb\Phi (t,\bar x_2, \bar x_3)=\beta  t-\frac{1}{2} \log \left(a_1^2+\kappa^2 \bar x_2^2 a_2^2+\left(e^{\bar x_3}-1\right)^2 a_3^2\right).
\end{equation}
Generalized Supergravity Equations with  $\wb{\mathcal J}=(0,0,0,0)$ are satisfied under the condition
\eqref{betapcond}.
Vectors  $\mathcal{\wh J}$, $\mathcal{\wb J}$ do
not agree with $Div\,\Theta$.

Transformation of Manin triple given by particular form of $C_2$
where
$$
c_{12}=c_{13}=c_{15}=c_{16}=0,\ c_{22}=c_{33}=1
$$
gives again rather extensive tensors and dilatons. The only relevant
difference is  that now $y_1= \hat x_1-\bar x_3$ and the Generalized
Supergravity Equations are satisfied with  $\mathcal{\wh J}=
(0,0,0,-1)$ and $\mathcal{\wb J}= (0,(\kappa +1) e^{-{\bar x_3}},0,0)$.
Vector  $\mathcal{\wb J}$ does
not agree with $Div\,\Theta$ that depends on the product $a_1(t)a_3(t)$.

\section{Conclusions}

We have presented many examples of \pltp y \tfn\ acting on flat or curved \bkg s invariant \wrt\ Bianchi groups ${\B}_{III}$, ${\B}_{V}$, ${\B}_{VI_{\kappa}}$ and ${\B}_{VI_{-1}}$. Coresponding dilatons were found using formulas \eqref{dualdil}, \eqref{phi0}.
In
many cases  the dilatons were non-local in the sense that they were both functions of coordinates $\hat x$ on $\hG$ and $\bar x$  on $\bG$. 
We have shown how to deal with this issue if the dependency
is linear. This partially resolves the puzzle explained in \ref{SUGRA}.

It turns out that plural \bkg s and dilatons often do not satisfy the usual beta function equations but Generalized Supergravity Equations provided the supplementary vector field $\mathcal{J}$ is computed using formulas \eqref{kilJ} or \eqref{kilJbar}  presented in Section \ref{SUGRA}. The formulas were repeatedly checked not only for the examples presented here, but also for other 
Manin triples given in \cite{snohla:ddoubles} and their embedings into $\cd$. All tested \bkg s and  dilatons  obtained by \PL\ \tfn s \eqref{Fhat}, \eqref{Fhat2}, \eqref{dualdil}, \eqref{phi0} satisfy Generalized Supergravity Equations.

As we noted in the Introduction, complete classification of plural models is beyond the scope of the paper as there are too many cases to discuss. Therefore, we present Manin triples that  demonstrate important properties of plural \sm s, dilatons and vectors $\mathcal{J}$. The examples show that vector fields $\mathcal J$ need not be constant as it turns out in cases of plurality $\mathfrak b_{III}\bowtie \mathfrak a$ to $\mathfrak b_{IIIiii}\bowtie \mathfrak b_{III}$, $\mathfrak b_{V}\bowtie \mathfrak a$ to $\mathfrak b_{Vii}\bowtie\mathfrak b_{VI_{-1}}$ and $\mathfrak b_{VI_\kappa}\bowtie\mathfrak a$ to $b_{VI_\kappa}\bowtie\mathfrak b_{VI_{-\kappa}.iii}$.
Let us note that  vector fields $\mathcal J$ are Killing vectors of corresponding \bkg s in spite of the fact that they are  linear combinations of left-invariant fields of corresponding groups that satisfy condition \eqref{kseq} with generally nonvanishing right-hand side.
Beside that, we have shown that the alternative formula
\eqref{Divtheta} for the supplementary vector $\mathcal J$ developed
for the \natd y does not work in general for \pltp y.

\end{document}